\documentclass[twocolumn,showpacs,preprintnumbers,amsmath,amssymb,prd,10pt,aps]{revtex4-1}

\usepackage{epsfig}
\usepackage{dcolumn}
\usepackage{bm}
\usepackage{comment}
\usepackage{amsmath,amssymb}

\usepackage{color}
\usepackage{graphicx}

\usepackage[utf8]{inputenc}

\newcommand{\be}{\begin{equation}}
\newcommand{\ee}{\end{equation}}
\newcommand{\ba}{\begin{eqnarray}}
\newcommand{\ea}{\end{eqnarray}}

\newcommand{\bea}{\begin{eqnarray*}}
\newcommand{\eea}{\end{eqnarray*}}

\graphicspath{{figures/}}

\begin{document}

\title{Extended Chameleons}

\author{Philippe Brax}
\affiliation{Institut de Physique Th\'eorique,\\
Universit\'e Paris-Saclay CEA, CNRS, F-91191 Gif-sur-Yvette, C\'edex, France\\}

\author{Nicola Tamanini}
\affiliation{Institut de Physique Th\'eorique,\\
Universit\'e Paris-Saclay CEA, CNRS, F-91191 Gif-sur-Yvette, C\'edex, France\\}

\begin{abstract}
We extend the  chameleon models by considering Scalar-Fluid theories where the coupling between matter and the scalar field can be represented by a quadratic effective potential with density-dependent minimum and mass. In this context, we study the
effects of the scalar field on Solar System tests of gravity and show that models passing these stringent constraints can still induce large modifications of Newton's law on galactic scales. On these scales we analyse models which could lead to a percent deviation of Newton's law outside the virial radius. We then model the dark matter halo as a Navarro-Frenk-White profile and explicitly find that the fifth force can give large contributions around the galactic core in a particular model where the scalar field mass is constant and the minimum of its potential varies linearly with the matter density. At cosmological distances, we find that this model does not alter the growth of large scale structures and therefore would be best tested on galactic scales, where interesting signatures might arise in the galaxy rotation curves.
\end{abstract}
\maketitle

\section{Introduction}
Scalar fields may play a fundamental role in the physics of the late time Universe \cite{Copeland:2006wr}. They could lead to the acceleration of the expansion of the Universe and/or modify gravity on large scales with
an impact on the growth of large scale structures \cite{Koyama:2015vza}. These effects have to be tamed by the requirement that local physics, i.e.~in the Solar System and the laboratory, does not deviate largely from
the behaviour anticipated for particles and objects described by standard physics. This can be achieved by essentially three known screening mechanisms: the chameleon \cite{Khoury:2003aq,Khoury:2003rn}, Vainshtein and K-mouflage mechanisms \cite{Joyce:2014kja}.
The Vainshtein \cite{Vainshtein:1972sx} and K-mouflage mechanisms \cite{Babichev:2009ee} imply that nearly spherical objects screen the scalar field within a certain (Vainshtein or K-mouflage) radius. In general, the Vainshtein mechanism for relevant astrophysical objects from stars to Galaxy clusters are screened, i.e.~they lie well within their Vainshtein radii. For K-mouflage, solar-like  stars are screened while galaxy clusters are not and galaxies are at the boundary between the screened and unscreened cases \cite{Brax:2014a}. For chameleon-like theories such as $f(R)$ in the large curvature regime, only objects with low surface Newton potentials can be unscreened, e.g.~dwarf galaxies \cite{Vikram:2013uba}.

Inverse power law chameleons are a prime example where the constraints from Solar System physics are stringent \cite{Khoury:2003rn}.
In this paper we will extend this class of chameleon theories and apply Solar System constraints to bound the parameters of these newly proposed models. There are three important tests. The first one is given by the absence of fifth force in the Solar System at the $10^{-5}$ level
as provided by the Cassini experiment \cite{Bertotti:2003rm}. The second one is the test of the strong equivalence principle in the Moon-Earth-Sun system at the $10^{-13}$ level given by the Lunar Ranging experiment \cite{Williams:2004qba,Williams:2012a}.
Finally the same experiment bounds the perihelion advance of the Moon around the Earth at the $10^{-11}$ level \cite{peri}. It turns out that the only mass scale of inverse power law  chameleon models is constrained to be less than $10^{-3}$ eV, a scale compatible with the dark energy scale now \cite{Khoury:2003aq,Khoury:2003rn}. These chameleon models can be mimicked in the Solar System using a density dependent massive scalar field whose potential has a minimum which also depends on the density of the environment.

In this paper, we generalise chameleon models to a generic scalar field model with a mass and a minimum which are both density-dependent and not related with each other. In other words the mass and the minimum of the scalar field potential can in principle be specified as completely arbitrary functions of the environmental density. We apply the three tests already mentioned in the Solar System and consider the behaviour of the scalar field on galactic scales. We focus on models which pass the Solar System tests and have a non-negligible effect on galaxies. In the galactic case, we first study the effects of the scalar interaction outside the virial radius and impose that the modifications of Newton's law should not exceed one percent. We then present an explicit integration of the scalar equation of motion for a Navarro-Frenk-White (NFW) profile \cite{Navarro:1996}, in the case where the mass is a constant and the minimum of the potential varies linearly with the density, and we find that the scalar force can have significant effects inside the galactic halo and in particular around the core size of the galaxy $R_s$. We also consider the physics on cosmological and galaxy cluster scales and in particular the growth of these structures in the linear regime.

The models considered in this paper can be embedded in the wider class of Scalar-Fluid theories \cite{part1,part2,Koivisto:2015qua,Boehmer:2015ina},
where completely general couplings between the scalar field and the matter sector (effectively described as a fluid) can be handled.
These theories have been recently introduced as a possible new framework for interacting dark energy where the field equations, being derived from a consistent Lagrangian, are well-defined at all orders, implying that not only the background cosmological equations \cite{part1,part2}, but also the perturbations equations \cite{Koivisto:2015qua,Boehmer:2015ina}, can be straightforwardly analysed.
Moreover Scalar-Fluid couplings appear to express new features which have previously been overlooked in the phenomenological construction of interacting dark energy models \cite{Tamanini:2015iia}.
Within such a framework the scalar-matter interaction can be investigated in full generality and extensions of well known models can be consistently formulated. As mentioned above, in this paper we will focus on extensions of chameleon theories. 
We will consider applications of these extended theories to scales ranging from the Solar System to cosmology.
In particular we will study the new screening mechanisms that can be obtained at Solar System and galactic distances, the possible effects on the galaxy rotation curves and finally the cosmological background and the growth of structures.

In more detail the plan of the paper is the following.
In Sec.~\ref{sec:theory} we will briefly review the theoretical setting of Scalar-Fluid theories, we will discuss the choice of the effective scalar-matter coupling and we will obtain the fifth force potential for a static non-relativistic environment.
Furthermore we will find general solutions for a constant sphere density profile, we will discuss the screening conditions and finally we will present the current Solar System constraints employed in this paper.
Using the results of Sec.~\ref{sec:theory}, in Sec.~\ref{sec:solar_system} we will then present a new class of screening models which naturally extend the chameleon paradigm.
Furthermore we will find the constraints on the parameters of different theories within this class, including also standard chameleons, given by the Solar System observations previously discussed.
In Sec.~\ref{sec:galaxy} we will investigate the possible effects of these new class of models at galactic scales, first outside the virial radius, assuming a constant density profile, and then inside the galactic halo, assuming a Navarro-Frenk-White (NFW) profile.
In the latter situation we will also show the interesting signatures that might arise in the galactic rotation curves.
In Sec.~\ref{sec:cosmo} we will then consider the dynamics of these new models at cosmological and cluster scales, analysing the possible deviations from $\Lambda$CDM in the growth of structures.
Finally we will discuss the results and draw our conclusions in Sec.~\ref{sec:conclusions}.

{\it Notation}. Throughout the whole paper we will use units with $\hbar = c = G =1$, measuring all physical quantities in GeV. We set $M_{\rm Pl} = 1/\kappa$ to be the (reduced) Planck mass.

\section{General theory and Solar System constraints}
\label{sec:theory}

In this first section we briefly introduce the general framework of Scalar-Fluid theories, focusing on the elements that will be important in the analysis that follows.
We will also provide the conditions under which the screening of the scalar field is effectively attained and will present the Solar System experiments that will be subsequently used to set constraints on the theoretical models.

\subsection{Scalar-Fluid action and scalar field equation}

The general action of Scalar-Fluid theories can be written as \cite{part1,Koivisto:2015qua}
\begin{align}
	\mathcal{S} = \int d^4x \left(\mathcal{L}_{\rm grav} +\mathcal{L}_{\rm fluid}+ \mathcal{L}_\phi+ \mathcal{L}_{\rm int}\right) \,,
	\label{eq:SF_action}
\end{align}
where $\mathcal{L}_{\rm fluid}$ represents matter in a perfect fluid form and is given by
\begin{align}
	\mathcal{L}_{\rm fluid} = -\sqrt{-g}\,\rho(n,s) + J^\mu\left(\varphi_{,\mu}+s\theta_{,\mu}+\beta_A\alpha^A_{,\mu}\right) \,,
	\label{001}
\end{align}
where $g$ is the determinant of the metric tensor $g_{\mu\nu}$ and $\rho$ is the energy density of the fluid. We assume $\rho(n,s)$ to be prescribed as a function of $n$, the particle number density, and $s$, the entropy density per particle.
$\varphi$, $\theta$ and $\beta_A$ are all Lagrange multipliers with $A$ taking the values $1,2,3$ and $\alpha_A$ are the Lagrangian coordinates of the fluid. The vector-density particle number flux $J^\mu$ is related to $n$ as
\begin{align}
	J^\mu=\sqrt{-g}\,n\,U^\mu\,, \quad |J|=\sqrt{-g_{\mu\nu}J^\mu J^\nu}\,, \quad n=\frac{|J|}{\sqrt{-g}} \,,
	\label{056}
\end{align}
where $U^\mu$ is the fluid 4-velocity satisfying $U_\mu U^\mu=-1$.
The details of the relativistic fluid description represented by the Lagrangian (\ref{001}) can be found in \cite{Brown:1992kc}.
The gravitational sector $\mathcal{L}_{\rm grav}$ is given by the standard Einstein-Hilbert Lagrangian
\begin{align}
	\mathcal{L}_{\rm grav} = \frac{\sqrt{-g}}{2\kappa^2}R \,,
\end{align}
where $R$ is the curvature scalar with respect to the metric $g_{\mu\nu}$, while the scalar field Lagrangian is given by
\begin{align}
	\mathcal{L}_\phi = -\sqrt{-g}\, \left[\frac{1}{2}\partial_\mu\phi\,\partial^\mu\phi +V(\phi)\right] \,,
\end{align}
with $V$ a general potential for $\phi$.
Finally for the interacting sector we will initially consider a general algebraic coupling of the type
\begin{align}
	\mathcal{L}_{\rm int} = -\sqrt{-g}\, f(n,s,\phi) \,, \label{027}
\end{align}
where $f(n,s,\phi)$ is an arbitrary function which will specify the particular model at hand.

Action~(\ref{eq:SF_action}) has to be varied with respect to the fields $g^{\mu\nu}$, $\phi$, $J^\mu$, $s$ and to the Lagrange multipliers $\varphi$, $\theta$, $\beta_A$, $\alpha^A$ (which are however not relevant in the analysis that follows).
The field equations in full detail are provided and discussed in \cite{part1,Koivisto:2015qua} and will not be repeated here.
We are only interested in the equation for the scalar field which reads
\begin{align}
	\Box\phi-\frac{\partial V}{\partial\phi}-\frac{\partial f}{\partial\phi} =0 \,,
	\label{eq:KG_covariant_eq}
\end{align}
where $\Box\phi = g^{\mu\nu} \nabla_\mu \nabla_\nu \phi$ with $\nabla_\mu$ the covariant derivative with respect to $g_{\mu\nu}$.
Note that the scalar field is subject to the effective potential
\begin{align}
  V_{\rm eff} = V(\phi) + f(n,s,\phi) \,,
  \label{eq:V_eff}
\end{align}
which generally depends on the distribution of surrounding matter through the particle number and entropy density of the fluid.

\subsection{Choice of the Scalar-Fluid coupling}
\label{sec:choice_of_coupling}

In our analysis we will make the two following simplifying assumptions:
\begin{itemize}
	\item The coupling function $f$ will not depend on the entropy density $s$ (adiabatic coupling).
	\item The fluid will describe only non-relativistic matter for which $\rho\propto n$. The scalar field will thus not be coupled to radiation which will be ignored in what follows.
\end{itemize}
Taken together these two requirements imply that the coupling function depends only on $\phi$ and $\rho$, i.e.~$f(\rho,\phi)$.

Note that the choice
\begin{equation}
	f=-\rho+\rho \exp(\beta\phi) \qquad\mbox{(standard chameleon)}  \,,
	\label{eq:chameleon}
\end{equation}
with $\beta$ constant, corresponds to the standard chameleon model \cite{Khoury:2003aq,Khoury:2003rn,Brax:2004qh}, as can be seen by looking at the resulting effective potential (\ref{eq:V_eff}) \footnote{The first term in Eq.~(\ref{eq:chameleon}) appears only to cancel the contribution coming from the fluid Lagrangian (\ref{001}). In chameleons theories in fact the whole non-relativistic matter sector is conformally coupled to the scalar field and no uncoupled term, beside radiation, is allowed.}.
The usual chameleon screening occurs since the height, mass (second derivative) and location of the minimum of the potential all change as $\rho$ changes.
All these quantities vary in a determined manner dictated by the specific form of the effective chameleon potential.
However in the Scalar-Fluid models we have the freedom to choose the function $f(\phi,\rho)$ at will and thus to let such quantities  vary independently.
In order to accomplish this, we can expand the effective potential (\ref{eq:V_eff}) around its minimum obtaining
\begin{equation}
	V_{\rm eff} (\rho,\phi) = V_0(\rho) + \frac{1}{2} m^2(\rho) \left[ \phi - \phi_0(\rho) \right]^2 + \dots \,,
	\label{eq:001}
\end{equation}
where $V_0$ is the value of $V_{\rm eff}$ at the minimum, $\phi_0$ its location and the mass is defined as
\begin{equation}
	m^2(\rho) = \left.\frac{\partial^2 V_{\rm eff}}{\partial^2\phi}\right|_{\phi_0} \,.
\end{equation}
All these quantities generally depends on $\rho$ and the dots in Eq.~(\ref{eq:001}) denote cubic and higher order terms in $\phi-\phi_0$.
Of course the linear order term in $\phi-\phi_0$ does not appear since we are expanding around the minimum of the potential.

The height of the minimum $V_0(\rho)$ is physically meaningless since it does not interact with the scalar field and can always be reabsorbed into the definition of the matter energy density.
To see this one can substitute $V_{\rm eff}$ back into the Scalar-Fluid action (\ref{eq:SF_action}).
Then one can redefine the matter energy density as $\rho \mapsto \tilde\rho = \rho + V_0(\rho)$ and thus require that the new function $\tilde\rho$ describes the physical energy density of non-relativistic matter.
Classically this correspond to the fact that the potential energy cannot be defined absolutely, but only its relative value carries a physical meaning.
Moreover a non vanishing $V_0$ would not affect the analysis that follows since its effects would not appear in Eq.~(\ref{eq:KG_covariant_eq}).

Given the considerations above in what follows we will assume an effective potential given by
\begin{equation}
	V_{\rm eff} (\rho,\phi) = \frac{1}{2} m^2(\rho) \left[ \phi - \phi_0(\rho) \right]^2 \,.
	\label{eq:eff_square_pot}
\end{equation}
This will well approximate any possible effective potential as long as the scalar field always satisfy $\vert \frac{\phi-\phi_0}{\phi_0} \vert\ll 1$, i.e.~as long as the scalar field takes values in the neighbourhood of the minimum $\phi_0$.

\subsection{Fifth force potential}

In Scalar-Fluid theories the geodesic motion of test particles is modified due to the presence of a fifth force $F_\mu$:
\begin{align}
	\frac{d U^\mu}{d\lambda}+\Gamma^\mu_{\sigma\nu} U^\sigma U^\nu  = F^\mu \,.
	\label{eq:geodesic_eq}
\end{align}
where $\Gamma^\mu_{\sigma\nu}$ is the Levi-Civita connection with respect to $g_{\mu\nu}$ and $\lambda$ is an affine parameter along the geodesic.
The general expression for $F_\mu$ has been derived in \cite{part1} and reads
\begin{equation}
	F^\mu = -\frac{h^{\nu\mu}}{\rho+p+p_{\rm int}+\rho_{\rm int}}\left(p_{,\nu}+p_{\rm int}{}_{,\nu}+\frac{\partial f}{\partial\phi}\phi_{,\nu}\right) \,,
	\label{eq:general_5th_force}
\end{equation}
where $p$ is the fluid pressure, the projective tensor $h_{\mu\nu}$ is defined as
\begin{equation}
	h_{\mu\nu} = g_{\mu\nu} + U_\mu U_\nu \,,
\end{equation}
and we used the general definitions
\begin{align}
	\rho_{\rm int} = f(n,s,\phi) \quad\mbox{and}\quad p_{\rm int} = n\frac{\partial f(n,s,\phi)}{\partial n}-f(n,s,\phi) \,.
\end{align}
For what concerns us here, given the assumptions outlined in Sec.~\ref{sec:choice_of_coupling}, the fifth force (\ref{eq:general_5th_force}) reduces to
\begin{equation}
	F^\mu =-h^{\nu\mu} \partial_\nu \left[\log \left( 1+ \frac{\partial f}{\partial \rho} \right) \right] \,.
\end{equation}

At this point we consider the Newtonian limit of the geodesic equation (\ref{eq:geodesic_eq}).
Within this limit the metric tensor can be taken to be diagonal with
\begin{equation}
	g_{00} = -1-2\Psi_N \quad\mbox{and}\quad g_{ij} = \delta_{ij}(1-2\Psi_N) \,.
\end{equation}
We will also assume that all velocities involved are negligible if compared with the speed of light and in particular we will consider
\begin{equation}
	U^\mu = \frac{\partial x^\mu}{\partial\lambda} = ( \frac{\partial t}{\partial\lambda}, 0, 0, 0 ) \,.
\end{equation}
Moreover all dynamical quantities will be assumed to be static, namely
\begin{equation}
	\frac{\partial \Psi_N}{\partial t} = 0 \,, \qquad \frac{\partial \phi}{\partial t}=0 \,, \qquad \frac{\partial\rho}{\partial t}=0 \,.
\end{equation}
Then the time-component of $F_\mu$ will vanish and the time-component of the geodesic equation (\ref{eq:geodesic_eq}) will reduce to $d^2t/d \lambda^2 = 0$ implying nothing but $d \lambda \propto dt$.
The spatial-components of the geodesic equation (\ref{eq:geodesic_eq}) will instead give
\begin{equation}
	\frac{d^2 x^i}{dt^2} + \frac{\partial \Psi_N}{\partial x^i} = - \frac{\partial}{\partial x^i} \left[\log \left( 1+ \frac{\partial f}{\partial \rho} \right) \right] \,,
\end{equation}
which can be rewritten as
\begin{equation}
	\frac{d^2 x^i}{dt^2} = - \frac{\partial}{\partial x^i} \left( \Psi_N + \Psi_5 \right) \,,
	\label{eq:Newton_2nd_law}
\end{equation}
where we have defined the {\it fifth force potential} as
\begin{equation}
	\Psi_5 = \log \left( 1+ \frac{\partial f}{\partial \rho} \right) \,.
\end{equation}
Eq.~(\ref{eq:Newton_2nd_law}) is nothing but Newton's second law of motion with an acceleration defined by a total potential given by the sum of the gravitational and fifth force potentials.
Wherever the fifth force potential is negligible in comparison with the Newtonian potential, test masses will not feel the fifth force and follow Newton's gravitational laws.
In other words the condition
\begin{equation}
	\Psi_5 \ll \Psi_N  \qquad\mbox{(screening condition)} \,,
\end{equation}
guarantees that all local effects of the fifth force have no physical consequences.

At this point we want to find the fifth force potential for the coupling function (\ref{eq:eff_square_pot}) that we will consider in the following sections.
First we note that in the standard chameleon case (\ref{eq:chameleon}) the fifth force potential coincides with the scalar field, namely~$\Psi_5 \propto \phi$.
In any other situation however the potential $\Psi_5$ will not be proportional to the scalar field and will in general depend on the matter energy density $\rho$.
If the scalar field effective potential (\ref{eq:eff_square_pot}) is chosen then the fifth force potential becomes
\begin{equation}
	\Psi_5 = \log \left[ 1 - m^2 \frac{\partial\phi_0}{\partial\rho} \left( \phi - \phi_0 \right) + \frac{1}{2} \frac{\partial m^2}{\partial\rho} \left( \phi - \phi_0 \right)^2 \right] \,.
	\label{eq:5th_force_pot}
\end{equation}
As mentioned above the effective potential (\ref{eq:eff_square_pot}) is expected to approximate any possible potential as long as $\vert \frac{\phi-\phi_0}{\phi_0} \vert \ll 1$.
In this situation the fifth force potential can be expanded  around the value of $\phi_0$.
The result is
\begin{multline}
	\Psi_5 = - m^2 \frac{\partial\phi_0}{\partial\rho} \left( \phi - \phi_0 \right) \\
	 + \frac{1}{2} \left[ \frac{\partial m^2}{\partial\rho} - m^4 \left( \frac{\partial\phi_0}{\partial\rho} \right)^2 \right] \left( \phi - \phi_0 \right)^2 + \dots \,,
\end{multline}
where dots denote again cubic and higher orders in $\phi - \phi_0$.

\subsection{Constant spherical profile}
\label{sec:sphere_profile}

Let us now consider a spherical distribution of matter with constant energy density $\rho_{\rm in}$ up to a certain radius $R$ and $\rho_{\rm out}$ outside such radius.
We will assume
\begin{align}
	\rho(r) &= \rho_{\rm in} \quad\mbox{for}\quad r\le R \,, \\
	\rho(r) &= \rho_{\rm out} \quad\mbox{for}\quad r> R \,,
\end{align}
where $r$ is the radial distance from the centre of the sphere.
Such  a matter distribution should ideally describe the Sun or a planet in the Solar System, with $\rho_{\rm in}$ the mean matter density inside the sphere and $\rho_{\rm out}$ the mean value for the interplanetary space within the Solar System.
In Sec.~\ref{sec:SS_constraints} we will use such idealization to put constraints on the parameters of our models.

An important issue must be made clear before proceeding.
The fifth force potential (\ref{eq:5th_force_pot}) depends not only on the scalar field, as it happens in chameleon theories, but it also depends on the matter energy density $\rho$.
This implies that if $\rho$ is discontinuous somewhere in  space-time, then an infinite fifth force will be present at that point providing thus a non physical result.
In the case of a constant sphere considered in this section, at the boundary of the sphere, i.e.~at $r=R$, the energy density $\rho$ is formally discontinuous and an infinite fifth force is expected.
Nevertheless such spherical profile should only be taken as an idealization valid only at scales sufficiently large so that the real solar or planetary matter distributions can effectively be described by a constant sphere.
For an investigation in the region $r\simeq R$ more suitable matter distributions should be employed.
If those matter distributions are continuous over the whole spacetime, as expected down to molecular and atomic scales, then no infinite fifth force will ever be present.
As a result we will match the profiles at $R(1- \epsilon)$ with the ones at $R(1+\epsilon)$, where $\epsilon \ll 1$ represents the width below which we cannot trust a discontinuous profile.
Using this procedure, we will have well-defined solutions for all $r$ where $\epsilon R \ll R$ is the cut-off scale of the model.

If now we assume the scalar field to depend only on the radial coordinate $r$, then equation (\ref{eq:KG_covariant_eq}) in spherical coordinate (for the Minkowski metric) becomes
\begin{equation}
	\phi''(r) + \frac{r}{2} \phi'(r) = m^2 \left[ \phi(r) - \phi_0 \right] \,,
	\label{eq:KG_eq_spherical}
\end{equation}
where a prime denotes differentiation with respect to $r$ and we recall that $m^2$ and $\phi_0$ depend on $\rho(r)$.
Our task now is to find two solutions of Eq.~(\ref{eq:KG_eq_spherical}): an inside solution $\phi_{\rm in}$ with $\rho=\rho_{\rm in}$ and an outside solution $\phi_{\rm out}$ with $\rho_{\rm out}$.
We will then match the two solutions requiring them to be continuous and differentiable at $r=R$, i.e.~requiring
\begin{equation}
 	\phi_{\rm in}(R) = \phi_{\rm out}(R) \quad\mbox{and}\quad \phi_{\rm in}'(R) = \phi_{\rm out}'(R) \,,
 	\label{eq:matching_sphere}
\end{equation}
and we will impose the following boundary conditions 
\begin{equation}
	\lim_{r\rightarrow 0} \phi_{\rm in}'(r) = 0 \quad\mbox{and}\quad \lim_{r\rightarrow \infty} \phi_{\rm out}(r) = \phi^{\rm out}_0 \,.
	\label{eq:bc_sphere}
\end{equation}
In what follows the definitions
\begin{align}
	\phi^{\rm out}_0 = \phi_0(\rho_{\rm out}) \,,\quad
	\phi^{\rm in}_0 = \phi_0(\rho_{\rm in}) \,, \\
	m^2_{\rm in} = m^2(\rho_{\rm in}) \,,\quad
	m^2_{\rm out} = m^2(\rho_{\rm out}) \,,
\end{align}
will be used in order to simplify the notation.

For both the inside and outside solutions $m^2$ and $\phi_0$ are nothing but constants in Eq.~(\ref{eq:KG_eq_spherical}).
In this case the general solution to Eq.~(\ref{eq:KG_eq_spherical}) can be written as
\begin{equation}
	\phi(r) = \phi_0 + \frac{c_1}{r} e^{m r} + \frac{c_2}{r} e^{-m r} \,,
\end{equation}
where $c_1$ and $c_2$ are two constants of integration.
Imposing the boundary conditions (\ref{eq:bc_sphere}) to this general solution one finds the inside and outside solutions as
\begin{align}
	\phi_{\rm in}(r) &= \phi_0^{\rm in} + \left(\phi_0^{\rm out}-\phi_0^{\rm in}\right) C_{\rm in} \frac{R}{r} \sinh \left( m_{\rm in} r \right) \,, \label{eq:sol_sphere_in}\\
	\phi_{\rm out}(r) &= \phi_0^{\rm out} + \left(\phi_0^{\rm in}-\phi_0^{\rm out}\right) C_{\rm out} \frac{R}{r} e^{m_{\rm out} (R-r)} \,, \label{eq:sol_sphere_out}
\end{align}
where $C_{\rm in}$ and $C_{\rm out}$ are dimensionless constants whose values is obtained computing the matching conditions (\ref{eq:matching_sphere}) as
\begin{align}
	C_{\rm in} &= \frac{\left(1+m_{\rm out}R\right)}{m_{\rm in}R\cosh\left(m_{\rm in}R\right)+m_{\rm out}R\sinh\left(m_{\rm in}R\right)} \,, \label{eq:const_in} \\
	C_{\rm out} &= \frac{m_{\rm in}R\cosh\left(m_{\rm in}R\right)-\sinh(m_{\rm in}R)}{m_{\rm in}R\cosh\left(m_{\rm in}R\right)+m_{\rm out}R\sinh\left(m_{\rm in}R\right)} \,. \label{eq:const_out}
\end{align}

At this point one can find the fifth force potential substituting the solutions (\ref{eq:sol_sphere_in}) and (\ref{eq:sol_sphere_out}) into Eq.~(\ref{eq:5th_force_pot}).
We will only consider the outside solution (\ref{eq:sol_sphere_out}), since the physical applications we are interested in concern only the region $r\gg R$.
Moreover we will focus on models where the screening might give rise to possible detectable effects, i.e.~we will neglect theories where the interaction between the scalar field and the surrounding matter is so small as to give a fifth force which is basically zero within the Solar System.
Such requirement can be translated in the following assumption:
\begin{itemize}
	\item $m_{\rm out} r_{\rm s.s.} \lesssim 1$ where $r_{\rm s.s.}$ can be taken as the maximal distance within the Solar System and estimated as $r_{\rm s.s.} \sim 10^{28}\ {\rm GeV}^{-1}$. This condition ensures that the exponential in (\ref{eq:sol_sphere_out}) does not become too small, forcing in this manner the outside solution to equal $\phi_0$ and thus the fifth force potential (\ref{eq:5th_force_pot}) to vanish everywhere (Yukawa suppression);
\end{itemize}
This assumption must be satisfied by any particular effective potential describing the Scalar-Fluid interaction, otherwise no observable effects of the fifth force will appear, at least at Solar System scales.
For each model we will refer to the part of its parameter space where this assumption is satisfied as the fifth force's ``detectable range'' in the Solar System.

Well within such range, i.e.~for $m_{\rm out} r_{\rm s.s.} \ll 1$, the fifth force potential can be taken to be
\begin{multline}
	\Psi_5 \simeq \log [ 1 - C_{\rm out}\, m^2_{\rm out} \left.\frac{\partial\phi_0}{\partial\rho}\right|_{\rm out} \left(\phi_0^{\rm in}-\phi_0^{\rm out}\right) \frac{R}{r}  \\
	+ \frac{C_{\rm out}^2}{2} \left.\frac{\partial m^2}{\partial\rho}\right|_{\rm out} \left(\phi_0^{\rm in}-\phi_0^{\rm out}\right)^2 \frac{R^2}{r^2}  ] \,,
	\label{eq:002}
\end{multline}
which at the second order in $R/r$ can be expanded as
\begin{multline}
	\Psi_5 \simeq - C_{\rm out} m^2_{\rm out} \left.\frac{\partial\phi_0}{\partial\rho}\right|_{\rm out} \left(\phi_0^{\rm in}-\phi_0^{\rm out}\right) \frac{R}{r} \\
	+ \frac{C_{\rm out}^2}{2} \left[\left.\frac{\partial m^2}{\partial\rho}\right|_{\rm out} -m^4_{\rm out} \left(\frac{\partial\phi_0}{\partial\rho}\right)_{\rm out}^2 \right] \left(\phi_0^{\rm in}-\phi_0^{\rm out}\right)^2 \frac{R^2}{r^2} \,,
	\label{eq:5th_force_pot_SS}
\end{multline}
where higher order terms in $R/r$ have been neglected.
This fifth force potential gives rise to corrections in $1/r$ and $1/r^2$ to the Newtonian potential, which can be constrained by Solar System observations; see Section~\ref{sec:SS_constraints}.

Before analysing the constraints given by Solar System tests, we first discuss some qualitative features that can be extracted from the fifth force potential just obtained.
From Eq.~(\ref{eq:5th_force_pot_SS}) it is clear that if the minimum of the effective potential does not change between inside and outside the sphere, namely if $\phi_0^{\rm in}=\phi_0^{\rm out}$, then $\Psi_5=0$ and no fifth force ever arises.
Thus for any signature of the scalar field to be detected within the Solar System, one must always have a screening mechanism where the minimum of the effective potential depends on the surrounding distribution of matter.
Note that this indeed happens in well known screening mechanisms such as the chameleon and symmetron mechanisms.
This is an interesting and important conclusion which tells us that deviations from Newton's law can appear only if the minimum of the effective potential depends on the environment.
In other words, if only the mass of the effective potential, and not its minimum, depends on the surrounding distribution of matter, then the fifth force will vanish everywhere, at least within the Solar System.

\subsection{Screening}

The models that we have introduced generalise the chameleon mechanism and the existence of a thin-shell. To illustrate this let us consider the case of a spherical object of radius $R_0$ and mass $M_0$. When $m_{\rm in }R_0\gg 1$ the field is nearly constant and equal to $\phi_0^{\rm in}$ inside the object and varies over a thin shell close to the surface. This is not the case when $m_{\rm in}R_0 \ll 1$ as the value of $\phi$ at the centre is equal to $\phi(0) = \phi_0^{\rm in} \sim \phi_0^{\rm out} $ and the overdensity due to the object has a small effect on the field profile. We can distinguish these two possibilities.

First we consider the case of a small object (ideally a test particle) of size $R_0$ such that $m_{\rm in} R_0 \sim m_{\rm out} R_0 \ll 1$ and for which $\phi(0) \sim \phi_0^{\rm out}$.
Within this approximation from Eq.~(\ref{eq:sol_sphere_out}) the field outside the object reads
\begin{equation}
\phi_{\rm out}(r) \simeq \phi_0^{\rm out} + \left.\frac{\partial\phi_0}{\partial\rho}\right|_{\rm out} m^2_{\rm out} \frac{M_0}{4\pi r} \,,
\end{equation}
and the potential equals
\begin{equation}
\Psi_5 (r) \simeq -\left(\frac{\partial\phi_0}{\partial\rho}\right)^2_{\rm out} m^4_{\rm out} \frac{M_0}{4\pi r} \,.
\end{equation}
For such objects behaving like test particles, this should be identified with
\begin{equation}
\left|\Psi_5 (r)\right| = 2 \beta_{\rm test}^2 \Phi_N(r) \,,
\end{equation}
where $\Phi_N$ is the Newtonian potential of the particle.
We  thus obtain
\begin{equation}
\beta_{\rm test}= M_{\rm Pl}\, m_{\rm out}^2 \left| \frac{\partial\phi_0}{\partial\rho}\right|_{\rm out} \,,
\label{eq:test_particle_coupling}
\end{equation}
with $M_{\rm Pl}$ the reduced Planck mass $M_{\rm Pl}^{-2} = 8\pi G_N$.
This measures the strength of the fifth force coupling between point particles.
Note that in general Eq.~(\ref{eq:test_particle_coupling}) depends on the surrounding distribution of matter $\rho_{\rm out}$, while in the standard chameleon one always has $\beta_{\rm test} =$ const.

Moreover Eq.~(\ref{eq:test_particle_coupling}) yields an extremely important result: if the minimum of the effective potential does not depend on the energy density, then $\beta_{\rm test}=0$ and consequently there is no fifth force interaction between two particles.
This is a general conclusion valid for any scalar field models irrespective of the form of its effective potential and in fact in well known screening mechanism, such as the chameleon and symmetron models, the minimum of the potential always changes with the environment.
Hence the constancy of the minimum precludes the appearance of the fifth force not only at the macroscopic level, i.e.~between extended spherical bodies (see Sec.~\ref{sec:sphere_profile}), but also at the microscopic one, namely between test particles.

We can now look at the case of objects extended in space, which cannot be idealized as test particles (this might be the case of the Moon orbiting Earth for example).
When the object is not such that $m_{\rm in} R_0 \ll1$, its fifth force potential is given by Eq.~(\ref{eq:5th_force_pot_SS}), which at first order in $R_0/r$ is
\begin{equation}
\Psi_5 (r)\simeq - m_{\rm out}^2 \frac{d\phi^{\rm out}_0}{d\rho} (\phi^{\rm in}_0 -\phi^{\rm out}_0) C_{\rm out} \frac{R_0}{r} \,.
\end{equation}
This can be identified with
\begin{equation}
\Psi_5 (r)= (1+ 2 \beta_{\rm object} \beta_{\rm test}) \Phi_N(r) \,,
\end{equation}
where $\beta_{\rm object}$ is the coupling strength of the object due to the fifth force.
Comparing these two  equations with Eq.~(\ref{eq:test_particle_coupling}) we obtain
\begin{equation}
\beta_{\rm object}= \frac{ \vert \phi^{\rm in}_0 -\phi^{\rm out}_0\vert }{2M_{\rm Pl}} \frac{C_{\rm out}}{\Phi_N(R_0)} \,.
\label{eq:beta_object}
\end{equation}
Such an object is screened if
\begin{equation}
\beta_{\rm object}\le \beta_{\rm test} \,.
\end{equation}
When $m_{\rm in}R \gg 1$, and consequently $C_{\rm out}\sim 1$, we recover the usual thin-shell condition of chameleon models \cite{Khoury:2003aq,Khoury:2003rn} here extended to a larger class of theories.
On the other hand when $m_{\rm in}R \ll 1$ and $m_{\rm in} \sim m_{\rm out}$ one can check that $\beta_{\rm object} \simeq \beta_{\rm test}$, in agreement with the previous analysis.

In the general case, these results can be extended to  interactions between screened bodies. Consider two bodies $A$ and $B$ with couplings $\beta_{A,B}$ as defined by (\ref{eq:beta_object}), then the potential felt by body $A$ and generated by body $B$ is simply
\begin{equation}
\Psi_5 (r)= (1+ 2 \beta_{A} \beta_{B}) \Phi_N(r) \,.
\end{equation}
We will use this result when calculating the perihelion of the moon in the vicinity of the earth. We can also calculate the effect of a diffuse distribution of matter creating the scalar field profile $\phi(\vec x)$ on a body of effective coupling $\beta_A$, the fifth force potential becomes
\be
\Psi_5 = \frac{\beta_A}{\beta_{\rm test}} \log (1+\frac{\partial f}{\partial \rho}) \,,
\ee
due to the screening effect $\beta_A\le \beta_{\rm test}$ on the coupled body $A$. This will be used for a dark matter halo acting on bodies moving either inside or outside the halo.

\subsection{Solar System constraints}
\label{sec:SS_constraints}

In this section we will use the fifth force potential (\ref{eq:5th_force_pot_SS}) to constrain the mass $m$ and minimum $\phi_0$ of the effective potential comparing possible predictions within the Solar Systems against the observational data.
Of course given a particular model, i.e.~given the functions $m^2(\rho)$ and $\phi_0(\rho)$ in terms of some parameters, the constraints derived in this way can also be used to put upper or lower bounds on the parameters.

In the fifth force potential (\ref{eq:5th_force_pot_SS}), the first term in $1/r$ gives a correction to Newton's law while the second term gives non-Newtonian corrections to planetary motions. At large enough distance $r\gg R$, these two terms are the leading effects to the orbits in the Solar System.
Using Eq.~(\ref{eq:5th_force_pot_SS}), it is convenient to define the ratio
\begin{multline}
	\epsilon\equiv \frac{\delta\Phi_N(r)}{\Phi_N(r)}= \frac{\Psi_5(r)}{\Phi_N(r)}\\
	= - C_{\rm out} m^2_{\rm out} \left.\frac{\partial\phi_0}{\partial\rho}\right|_{\rm out} \left(\phi_0^{\rm in}-\phi_0^{\rm out}\right) \frac{R}{G_N M}+\dots \,,
	\label{eq:eps_general}
\end{multline}
where $\delta \Phi_N = \Psi_5$ is the fifth force correction to Newton's potential $G_NM/r$ and $M$ is the mass of the objects around which other test objects orbit, i.e.~either the Sun for planets or the Earth for the Moon. The dots represent the higher order corrections in $R/r$ which can be neglected when $r\gg R$.
This ratio is constrained at the $10^{-5}$ level in the Solar System when $M= M_\odot$ for the sun-planets system \cite{Bertotti:2003rm}, namely
\begin{equation}
	\left|\epsilon_\odot\right| \approx C_\odot^{\rm s.s.} m^2_{\rm s.s.} \left.\frac{\partial\phi_0}{\partial\rho}\right|_{\rm s.s.} \left(\phi_0^{\odot}-\phi_0^{\rm s.s.}\right) \frac{1}{\Phi_\odot} \lesssim 10^{-5} \,,
	\label{constr_eps}
\end{equation}
where we have defined $\Phi_\odot\equiv \Phi_N (R_\odot) \sim 10^{-6}$ to be the Newtonian potential on the Sun's surface and where the outside matter density is taken as the mean interplanetary density, coinciding roughly with the mean galactic density $\rho_{\rm s.s.} \sim 10^{-41} \ {\rm GeV}^4$.
We have also defined $C^{\rm s.s.}_{\odot}$ to be $C_{\rm out}$ evaluated at $\rho_{\rm in} = \rho_\odot$, $\rho_{\rm out} = \rho_{\rm s.s.}$ and $R = R_\odot$.

On the other hand, the perihelion advance due to the non-Newtonian force in $1/r^3$ is given by
\begin{equation}
	\delta\varphi= \pi r \frac{d}{dr}\left[ r^2 \frac{d}{dr}\left(\frac{\epsilon}{r}\right)\right] \,,
\end{equation}
and thanks to Eq.~(\ref{eq:5th_force_pot_SS}) it is dominated by
\begin{multline}
	\delta\varphi= \pi\, C_{\rm out}^2 \left[\left.\frac{\partial m^2}{\partial\rho}\right|_{\rm out} -m^4_{\rm out} \left(\frac{\partial\phi_0}{\partial\rho}\right)_{\rm out}^2 \right] \\
	\times \left(\phi_0^{\rm in}-\phi_0^{\rm out}\right)^2 \frac{R^2}{G_N M r} \,.
\end{multline}
which can be rewritten as
\begin{equation}
	\delta\varphi= \pi\, C_{\rm out}^2 \left.\frac{\partial m^2}{\partial\rho}\right|_{\rm out}  \left(\phi_0^{\rm in}-\phi_0^{\rm out}\right)^2 \frac{R^2}{G_N M r} -\epsilon^2 \frac{G_N M}{r} \,,
	\label{eq:003}
\end{equation}
In the Earth-Moon system, this is constrained at the $10^{-11}$ level \cite{peri}.
Using $\epsilon\lesssim 10^{-5}$ and the Newtonian potential of all objects in the Solar System $G_N M/r \lesssim \Phi_{\odot}=10^{-6}$, we see that the second term in Eq.~(\ref{eq:003}) is negligible compared to the constraint on $\delta\varphi$ at the $10^{-11}$ level.
In the Earth-Moon system we impose therefore
\begin{multline}
	\delta\varphi_{\rm Moon} \approx \\
	\pi\, (C_\oplus^{\rm s.s})^2 \left.\frac{\partial m^2}{\partial\rho}\right|_{\rm s.s.}  \left(\phi_0^{\oplus}-\phi_0^{s.s.}\right)^2 \frac{1}{\Phi_\oplus}\frac{R_\oplus}{ r_{\rm Moon}} \lesssim 10^{-11} \,,
	\label{constr_phi}
\end{multline}
where the Newtonian potential on the Earth's surface can be estimated as $\Phi_\oplus \sim 10^{-9}$ and $R_\oplus$ and $r_{\rm Moon}$ are respectively the Earth's radius and the Moon-Earth distance.
Again we have defined $C^{\rm s.s.}_{\oplus}$ to be $C_{\rm out}$ evaluated at $\rho_{\rm in} = \rho_\oplus$, $\rho_{\rm out} = \rho_{\rm s.s.}$ and $R = R_\oplus$.

This last result only applies when the Moon is not screened and behaves like a test particle in the gravitational field of the Earth. When this is not the case, i.e.~when $\beta_{\rm Moon} < \beta_{\rm test}$, the perihelion constraint is relaxed.
For this situation we can take into account the Moon screening effect multiplying the previous result by the ratio $\beta_{\rm Moon}/\beta_{\rm test}$, namely
\begin{equation}
	\hat{\delta\varphi}_{\rm Moon} = \frac{\beta_{\rm Moon}}{\beta_{\rm test}} \delta\varphi_{\rm Moon}  \lesssim 10^{-11} \,,
	\label{constr_phi_screened}
\end{equation}
which should provide a more accurate constraint.
When the Moon is not screened, i.e.~when $\beta_{\rm Moon} \simeq \beta_{\rm test}$, one recovers the expression in Eq.~(\ref{constr_phi}).

Finally, the test of the equivalence principle in the Earth-Moon system restricts the E\"otvos parameter as \cite{Williams:2012a}
\begin{equation}
\eta_{\rm Moon-Earth}= \frac{\vert a_{\rm Moon}-a_{\rm Earth}\vert}{2(a_{\rm Moon}+a_{\rm Earth})}\le 10^{-13} \,,
\end{equation}
where $a_{\rm Moon}$ and $a_{\rm Earth}$ are the relative acceleration of the Moon and the Earth to the Sun.
Measuring these quantities leads to a constraint on the coupling of the Earth to the scalar field \cite{Khoury:2003rn}
\begin{equation}
\beta_{\rm Earth} = \frac{ \vert \phi^{\rm Earth}_0 -\phi^{\rm s.s}_0\vert }{2M_{\rm Pl}} \frac{C_\oplus^{\rm s.s.}}{\Phi_\oplus}  \le 10^{-7} \,.
\label{constr_eta}
\end{equation}

To summarise, we can thus use these three tests in the Solar System, namely Eqs.~(\ref{constr_eps}), (\ref{constr_phi_screened}) and (\ref{constr_eta}), to constrain our Scalar-Fluid models.
In practice given a particular model in terms of some parameters, i.e.~given the functions $\phi_0(\rho)$ and $m^2(\rho)$, these constraints can be used to restrict the allowed parameter space of the model.
An interesting example is provided by the models considered in the next sections, which nicely extend the standard chameleon paradigm.

\section{Extended chameleons at Solar System scales}
\label{sec:solar_system}

One can apply the Solar System constraints derived in the previous section to a specific model where the functions $\phi_0(\rho)$ and $m^2(\rho)$ have been specified.
There are no restrictions on the types of functions one can use for $\phi_0(\rho)$ and $m^2(\rho)$ and in principle any dependence might be assumed.
We will however focus the following analysis on power-law functions since, as we are going to see, they include and extend the well-known chameleon screening.

\subsection{Extending the chameleon model}

In the inverse power-law chameleon screening the effective potential of the scalar field reads
\begin{equation}
	V_{\rm cham}(\phi) = \frac{M_c^{4+\alpha}}{\phi^\alpha} + \rho\, e^{\beta \phi/M_{\rm Pl}} \,,
	\label{chameleon_pot}
\end{equation}
where $\alpha$ and $\beta$ are two positive and dimensionless parameters, $M_{\rm Pl} \sim 10^{18}\ {\rm GeV}$ is the (reduced) Planck mass and $M_c$ is a mass scale roughly constrained to be less than $10^{-12}\ {\rm GeV}$ by Solar System observations \cite{Khoury:2003rn}.
Given the physical requirements $\rho \ll M_{\rm Pl}^4$ and $\phi \ll M_{\rm Pl}$, the minimum and mass of the potential (\ref{chameleon_pot}) can be expressed as
\begin{align}
	\phi_0^{\rm cham} &\simeq M_c \left( \frac{\beta}{\alpha} \frac{\rho}{M_c^3 M_{\rm Pl}} \right)^{-\frac{1}{\alpha+1}} \,, \label{minimum_chameleon} \\
	m^2_{\rm cham} &\simeq M_c^2\, \alpha (\alpha + 1) \left( \frac{\beta}{\alpha} \frac{\rho}{M_c^3 M_{\rm Pl}} \right)^{\frac{\alpha+2}{\alpha+1}} \,. \label{mass_chameleon}
\end{align}
For the chameleon mechanism the minimum and mass of the effective potential are thus effectively provided by power-law functions of $\rho$.

In light of this result we will choose a Scalar-Fluid model where the minimum and mass of the effective potential are given by
\begin{align}
	\phi_0 &= M_1 \left( \frac{\rho}{M_1^3 M_{\rm Pl}} \right)^{-p} \,, \label{power_law_minimum} \\
	m^2 &= M^2_2 \left( \frac{\rho}{M_2^3 M_{\rm Pl}} \right)^{n} \,, \label{power_law_mass}
\end{align}
where $p$ and $n$ are now free parameters and $M_1$, $M_2$ are new mass scales which we are going to constrain using Solar System observations.
The chameleon case is easily recovered setting
\begin{equation}
	p = \frac{1}{\alpha+1} \quad \mbox{and} \quad n = \frac{\alpha+2}{\alpha+1} \,,
	\label{eq:cham_parameters}
\end{equation}
in which case the powers are also restricted to the ranges $0<p<1$ and $1<n<2$, and relating the mass scales as
\begin{align}
	M_1 &= M_c\, \alpha ^{\frac{1}{\alpha +4}} \beta ^{-\frac{1}{\alpha +4}} \,, \label{eq:cham_M1} \\
	M_2 &= M_c\, \alpha ^{\frac{1}{\alpha +4}} (\alpha +1)^{-\frac{\alpha +1}{\alpha +4}} \beta ^{-\frac{\alpha +2}{\alpha +4}} \,. \label{eq:cham_M2}
\end{align}
Notice that the chameleon parameter $\beta$ coincides with
\be
\beta\equiv \beta_{\rm test} = {\rm const} \,,
\ee
as obtained for Scalar-Fluid models substituting the relations (\ref{eq:cham_parameters})--(\ref{eq:cham_M2}) into Eq.~(\ref{eq:test_particle_coupling}).
This implies that from the extended model (\ref{power_law_minimum})--(\ref{power_law_mass}) one can truly recover standard chameleons setting the parameters as in Eqs.~(\ref{eq:cham_parameters})--(\ref{eq:cham_M2}).

The generalised model (\ref{power_law_minimum})--(\ref{power_law_mass}) thus extends the chameleon screening letting the minimum and mass  depend on unrelated powers of $\rho$ and breaking the relation between the two mass scales.
Note that $n=0$ and $p=0$ respectively correspond to the cases where the mass and the minimum do not depend on the matter density,  the latter being unconstrained within the Solar System (as discussed at the end of Sec.~\ref{sec:sphere_profile}).
With the Scalar-Fluid model (\ref{power_law_minimum})--(\ref{power_law_mass}) we will thus be able to test the unexplored scenario where $p$, $n$ and $M_1$, $M_2$ are unrelated and lie outside their corresponding chameleon ranges defined by $M_c$ and $\alpha$.

\subsection{Screening}

It is well-known that in the standard chameleon mechanism the coupling strength between two test particles is constant and coincide with the parameter $\beta$, which determines nothing but the coupling between a test particle and the scalar field.
For the extended models just introduced one must however use Eq.~(\ref{eq:test_particle_coupling}) to evaluate the fifth force strength between two test particles.
This generally provide an environment depending coupling strength which, using the extended chameleon model (\ref{power_law_minimum})--(\ref{power_law_mass}), reads
\begin{equation}
	\beta_{\rm test}^{\rm EC} = |p| M_1^{1+3p} M_2^{2-3n} \left( \frac{\rho_{\rm out}}{M_{\rm Pl}} \right)^{n-p-1} \,.
	\label{eq:beta_test_EC}
\end{equation}
As we mentioned above, using Eqs.~(\ref{eq:cham_parameters})--(\ref{eq:cham_M2}) one can check that for standard chameleon models $n-p-1=0$ and $\beta_{\rm test}= \beta$ is retrieved.
This implies that the model (\ref{power_law_minimum})--(\ref{power_law_mass}) is indeed a direct extension of the chameleon mechanism where the coupling strength between two point particles is allowed to depend on the local environment.
Note however that the chameleon model is not the only one where $\beta_{\rm test}$ is constant: in principle any model for which $n-p-1=0$, e.g.~$n=2$ and $p=1$, will possess such features.
In such cases the coupling strength $\beta_{\rm test}$ will be determined by the ratio between the two mass scales $M_1$ and $M_2$.
Moreover note that if $p=0$ we have $\beta_{\rm test}=0$, meaning that there is no fifth force at all between two test particles.
This result is in agreement with the considerations we have made above, when we have shown that the fifth force is completely screened if the minimum of the potential does not change.
On the other hand a constant scalar field mass, corresponding to $n=0$, allows for a non-vanishing force between two test particles.
If in addition $p=-1$, corresponding to a minimum linearly varying in $\rho$, then the resulting $\beta_{\rm test}$ is constant, exactly as in the standard chameleon case.
This specific model will be particularly useful in Sec.~\ref{sec:NFW}, where we will analyse the effects of extended chameleons on galaxy rotation curves, and it will be further discussed in Sec.~\ref{sec:conclusions}.

We can also compute the screening of an object of finite dimensions as given by Eq.~(\ref{eq:beta_object}).
In this case we find
\begin{equation}
	\beta^{\rm EC}_{\rm object}= \frac{C_{\rm out}}{\Phi_N(R_0)} \frac{M_1}{2M_{\rm Pl}} \left(\frac{\rho_{\rm out}}{M_1^3 M_{\rm Pl}}\right)^{-p} \left| \left(\frac{\rho_{\rm in}}{\rho_{\rm out}}\right)^{-p} -1 \right|  \,,
	\label{eq:beta_object_EC}
\end{equation}
where $\rho_{\rm in}$ is the density inside the object, e.g.~the mean density of the Moon, and $C_{\rm out}$ must of course be evaluated with the model (\ref{power_law_minimum})--(\ref{power_law_mass}).
In the situation where $C_{\rm out} \ll 1$, usually happening when $m_{\rm in}R \gg 1$, the object is screened.
Note that the $p=0$ case gives $\beta_{\rm object}=0$, again stressing the fact that no physical effects of the fifth force are present when the minimum does not depend on the local energy density.

\subsection{Solar System constraints}

At this point we want to use the Solar System observations given by Eqs.~(\ref{constr_eps}), (\ref{constr_phi_screened}) and (\ref{constr_eta}) to constrain the parameters $p$ and $n$ and the mass scales $M_1$ and $M_2$ of the extended chameleon model (\ref{power_law_minimum})--(\ref{power_law_mass}).
The first thing we need to compute however is the Solar System detectable range of the model (\ref{power_law_minimum})--(\ref{power_law_mass}) given by the requirement $m_{\rm out} r_{\rm s.s.} \lesssim 1$ (no Yukawa suppression).
Within this range the fifth force can be derived by the potential (\ref{eq:5th_force_pot_SS}) with good accuracy.
For this evaluation the outside matter density is assumed to be $\rho_{s.s.}\sim 10^{-41}\ {\rm GeV}^4$, which approximately corresponds to the mean galactic density, while the maximum distance within the Solar System is estimated as $r_{\rm s.s.} \sim 10^{28}\ {\rm GeV}^{-1}$, roughly the distance between Pluto and the Sun.
After having determined the detectable range in this way, we can use the Solar System constraints just mentioned to deduce what part of it is actually forbidden by the observations.
In other words we will determine what part of the parameter space can be excluded by Solar System tests.

Let us now see how the constraints (\ref{constr_eps}), (\ref{constr_phi_screened}) and (\ref{constr_eta}) can be implemented for the model (\ref{power_law_minimum})--(\ref{power_law_mass}).
In this case the detectable range is determined by the mass scale $M_2$ and parameter $n$, since in the condition $m_{\rm out} r_{\rm s.s.} \lesssim 1$ only the mass of the effective potential appears:
\begin{equation}
	 M_2 \left( \frac{\rho_{\rm s.s.}}{M_2^3 M_{\rm Pl}} \right)^{\frac{n}{2}} r_{\rm s.s.} \lesssim 1 \,.
	 \label{SS_det_range}
\end{equation}
However in the Solar System constraints  on the mass scale $M_1$ and the parameter $p$ also appear.
In fact using the extended models (\ref{power_law_minimum})--(\ref{power_law_mass}), the first constraint (\ref{constr_eps}) becomes
\begin{multline}
	\left|\epsilon_\odot\right| \approx C_\odot^{\rm s.s.} \frac{|p|}{\Phi_\odot} \frac{M_1^2 M_2^2}{\rho_{\rm s.s.}} \left( \frac{\rho_{\rm s.s.}}{M^3_2 M_{\rm Pl}} \right)^n \\
	\times \left( \frac{\rho_{\rm s.s.}}{M^3_1 M_{\rm Pl}} \right)^{-2p} \left| \left( \frac{\rho_{\rm s.s.}}{\rho_\odot} \right)^{p} -1 \right| \lesssim 10^{-5} \,,
	\label{SS_constr_1}
\end{multline}
while the second one (\ref{constr_phi}), generally true even if the Moon is screened with $\beta_{\rm Moon}< \beta_{\rm test}$, yields
\begin{multline}
	\hat{\delta\phi}_{\rm Moon} \approx \frac{\beta_{\rm Moon}}{\beta_{\rm test}} (C_\oplus^{\rm s.s})^2 \frac{\pi |n|}{\Phi_\oplus} \frac{M^2_1 M^2_2}{\rho_{\rm s.s.}} \left( \frac{\rho_{\rm s.s.}}{M^3_2 M_{\rm Pl}} \right)^n \\
	\times \left( \frac{\rho_{\rm s.s.}}{M^3_1 M_{\rm Pl}} \right)^{-2p} \frac{R_\oplus}{r_{\rm Moon}} \left[ \left( \frac{\rho_{\rm s.s.}}{\rho_\oplus} \right)^{p} -1 \right]^2 \lesssim 10^{-11} \,,
	\label{SS_constr_2bis}
\end{multline}
where we recall that the test particle expression (\ref{constr_phi}) has been suppressed by the factor $\beta_{\rm Moon}/\beta_{\rm test}$, which in the present case is determined by Eqs.~(\ref{eq:beta_test_EC}) and (\ref{eq:beta_object_EC}).
Finally the third constraint (\ref{constr_eta}) becomes
\begin{equation}
	\beta^{\rm EC}_{\rm object}= \frac{C^{\rm s.s.}_\oplus}{\Phi_\oplus} \frac{M_1}{2M_{\rm Pl}} \left(\frac{\rho_{\rm s.s.}}{M_1^3 M_{\rm Pl}}\right)^{-p} \left| \left(\frac{\rho_{\rm s.s.}}{\rho_\oplus}\right)^{p} -1 \right| \leq 10^{-7} \,.
	\label{SS_constr_3}
\end{equation}
In all these constraints the only free parameters are $p$, $n$, $M_1$ and $M_2$, while any other quantities are physically determined: $\Phi_\odot \simeq 4\times 10^{-6}$ (Newtonian potential on the Sun's surface), $\Phi_\oplus \simeq 7.2\times 10^{-10}$ (Newtonian potential on the Earth's surface), $\Phi_{\rm Moon} \simeq 3.3\times 10^{-11}$ (Newtonian potential on the Moon's surface), $\rho_{s.s.} \sim 10^{-41} {\rm GeV}^4$ (mean interplanetary matter density), $\rho_\odot \simeq 6.2\times 10^{-21} {\rm GeV}^4$ (mean density of the Sun), $\rho_\oplus \simeq 2.4\times 10^{-20} {\rm GeV}^4$ (mean density of the Earth), $\rho_{\rm Moon} \simeq 1.5\times 10^{-20} {\rm GeV}^4$ (mean density of the Moon), $R_{\rm Moon} \simeq 8.6\times 10^{21} {\rm GeV}^{-1}$ (radius of the Moon) and $R_\oplus/r_{\rm Moon} \simeq 6\times 10^{-3}$ (ratio between Earth's radius and Earth-Moon distance).
Moreover $C^{\rm s.s.}_\oplus$ and $C^{\rm s.s.}_\odot$ are given by Eq.~(\ref{eq:const_out}) with $\rho_{\rm out} = \rho_{\rm s.s}$ and, respectively, $\rho_{\rm in} = \rho_\oplus$ plus $R = R_\oplus \simeq 3.2\times 10^{22} {\rm GeV}^{-1}$ and $\rho_{\rm in} = \rho_\odot$ plus $R = R_\odot \simeq 3.5\times 10^{24} {\rm GeV}^{-1}$.

\subsubsection{Standard chameleons}

\begin{figure}
\includegraphics[width=0.95\columnwidth]{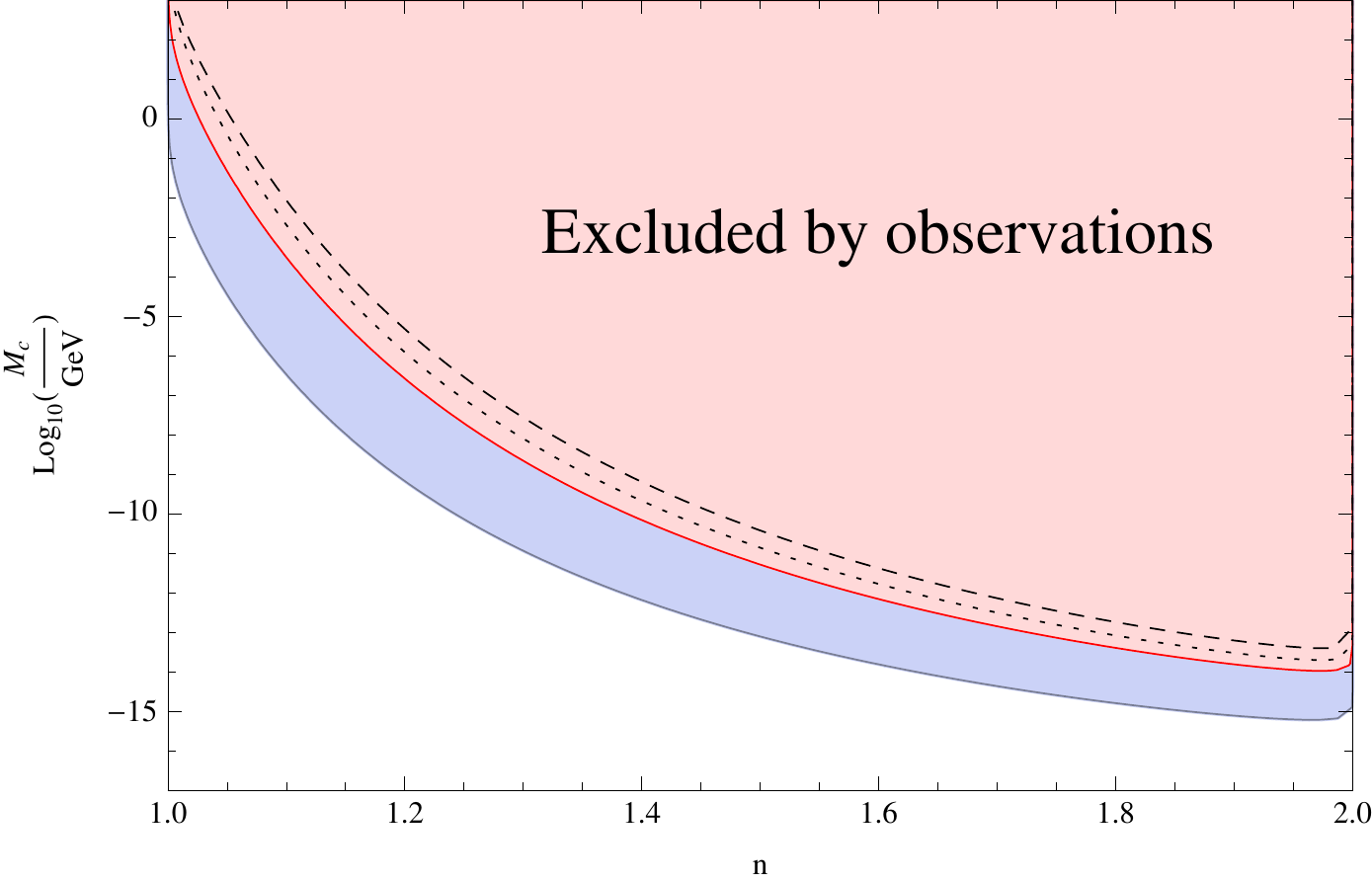}
\caption{Solar System constraints for the chameleon model (\ref{minimum_chameleon})--(\ref{mass_chameleon}). The red zone is forbidden by observations, the blue zone is the allowed part of the detectable range given by condition (\ref{SS_det_range}), while the rest of the parameter space (white region) is viable due to the Yukawa suppression. The stronger constraint (limit of the red region) is given by Eq.~(\ref{SS_constr_1}), while the dashed and dotted lines denote the constraints (\ref{SS_constr_2bis}) and (\ref{SS_constr_3}), respectively.}
\label{Fig:det_range_chameleon}
\end{figure}
Before  analysing the full Scalar-Fluid model (\ref{power_law_minimum})--(\ref{power_law_mass}), we will show what conclusions can be derived for the chameleon case, where the minimum and mass of the effective potential are given by (\ref{minimum_chameleon})--(\ref{mass_chameleon}).
Assuming a matter to scalar coupling of order one, $\beta\sim 1$, the detectable range and its corresponding zone excluded by Solar System constraint have been drawn in Fig.~\ref{Fig:det_range_chameleon}.
Recall that the value of $n$ is related to $\alpha$ by Eq.~(\ref{eq:cham_parameters}), and thus is constrained to be between 1 and 2 for $0<\alpha<+\infty$.
The dominant constraint for chameleons is (\ref{SS_constr_1}), which delimit the excluded (red) region in Fig.~\ref{Fig:det_range_chameleon}.
The dashed and dotted lines denote instead the constraints coming from Eqs.~(\ref{SS_constr_2bis}) and (\ref{SS_constr_3}), respectively.
Below the dashed line the Moon is effectively screened ($\beta_{\rm Moon}/\beta_{\rm test} \lesssim 10^{-4}$).
However if the Moon were not screened then the constraint (\ref{SS_constr_2bis}) would actually be the stronger one and even more detectable range would be excluded by the observations.
Higher values of the chameleon mass scale ($M_c\gtrsim 10^{5}$) are excluded by both Eqs.~(\ref{SS_constr_1}) and (\ref{SS_constr_3}), while the condition (\ref{SS_constr_2bis}) fails to constrain this part of the parameter space.

Note how the observations exclude a great part of the detectable range, although some of it is still available and could in principle be probed by future experiments.
Fig.~\ref{Fig:det_range_chameleon}  summarises the present phenomenological situation for the chameleon model (\ref{minimum_chameleon})--(\ref{mass_chameleon}). Our results complement the bounds already given in \cite{Khoury:2003rn}.


\subsubsection{Single mass scale models}

\begin{figure*}
\includegraphics[width=0.9\textwidth]{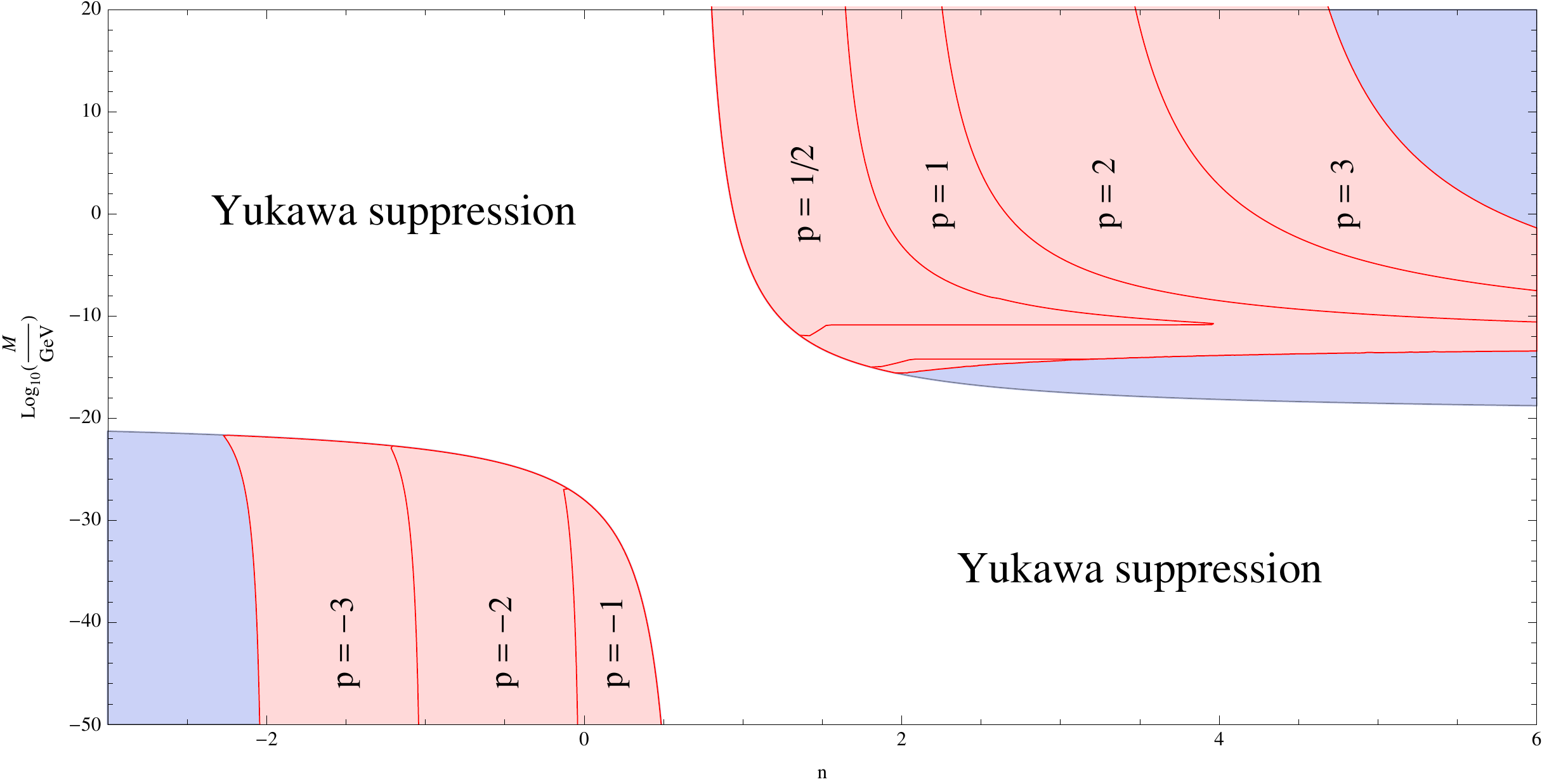}
\caption{Detectable range and Solar System constraints for model (\ref{constr_eps}) and (\ref{constr_phi}) with $M_1 = M_2 = M$. Red regions are excluded by the experiments, blue regions denote Solar System the detectable range, while in the white zone the fifth force is Yukawa suppressed and consequently undetectable.}
\label{Fig:det_range_equal_mass}
\end{figure*}
In order to simplify the analysis of the extended models (\ref{power_law_minimum})--(\ref{power_law_mass}), we will first consider the case where the two mass scales coincide, namely $M_1 = M_2 = M$.
For this case the detectable range, still given by Eq.~(\ref{SS_det_range}), and the corresponding Solar System constraints have been depicted in Fig.~\ref{Fig:det_range_equal_mass} for both positive ($p= 1/2, 1, 2, 3$) and negative ($p= -1, -2, -3$) values of $p$.
The region of the detectable range forbidden by observations is shown in red for the chosen values of $p$.
In the figure the regions excluded by higher absolute values of $p$ include the ones denoted with corresponding lower absolute values. For example the region excluded by $p=2$ includes also the areas denoted with $p=1$ and $p=0.5$ and analogously the region excluded by say $p=-3$ include the ones denoted by $p= -1$ and $p=-2$ as well.
Note also that in the bottom left of the whole positive $p$ red region there is a small area which is not excluded by $p=1$, but which is excluded by $p=2$ and by higher values of $p$.
The white region in Fig.~\ref{Fig:det_range_equal_mass} corresponds again to parameters choices for which the fifth force is Yukawa suppressed, and thus physically viable since its effects will never be strong enough in the Solar System.

Interestingly the detectable range now extends also to negative and vanishing values of $n$, although there high values of $M$ are Yukawa suppressed while lower values are potentially detectable.
This is an expected feature considering that $m^2$ is directly proportional to $M$ for $n<0$, and thus the higher the value of $M$ the more suppressed the Yukawa potential will be, the opposite trend happening for $n>0$.
Models with a constant effective potential mass, namely $n=0$, allow for possible values of $M$ to be within the detectable range, as Fig.~\ref{Fig:det_range_equal_mass} shows.
This is an interesting feature suggesting that models where only the minimum of the effective potential depends on the matter density can actually be constrained by Solar System observations.

The dominant constraints in Fig.~\ref{Fig:det_range_equal_mass} are the equivalence principle bounds given by Eq.~(\ref{SS_constr_3}) and the Cassini bound given by Eq.~(\ref{SS_constr_1}).
The first one forbids all parts of the red regions for $n\gtrsim 0.8$, i.e.~for positive $p$, except the little ``protuberances'' appearing in the bottom left of $p=1/2$ and $p=1$ which are actually excluded by the Cassini constraint (\ref{SS_constr_1}).
The $n\lesssim 0.8$ constraints are instead dominated by the Cassini experiment (\ref{SS_constr_1}), although the constraint (\ref{SS_constr_3}) gives comparable bounds.
From Fig.~\ref{Fig:det_range_equal_mass} it is clear that the higher the absolute value of $p$, the more of the parameter space is excluded by observations.
However while the limit $p\rightarrow -\infty$ manages to exclude the whole bottom left part of the detectable range, the limit $p\rightarrow +\infty$ fails to exclude the whole top right part.
In fact the low mass region of the $n\gtrsim 0.8$ detectable range, roughly between $M \sim 10^{-20}$ and $M \sim 10^{-13}$, under the excluded red zone, is never forbidden by observations, irrespectively of the value of $p$.
Nevertheless high values of $n$ instead are excluded by models with $p>1$ in a small strip of region roughly above $M \simeq 10^{-13} {\rm GeV}$.
For this feature the higher the value of $p$, the wider  the strip and the higher the values of $n$ that will be forbidden.
On the other hand in the limit $p\rightarrow 0$ basically the entire detectable range becomes viable, again in agreement with the fact that no fifth force effects are present if the minimum of the effective potential does not depend on the matter density.

\begin{figure}
\includegraphics[width=0.95\columnwidth]{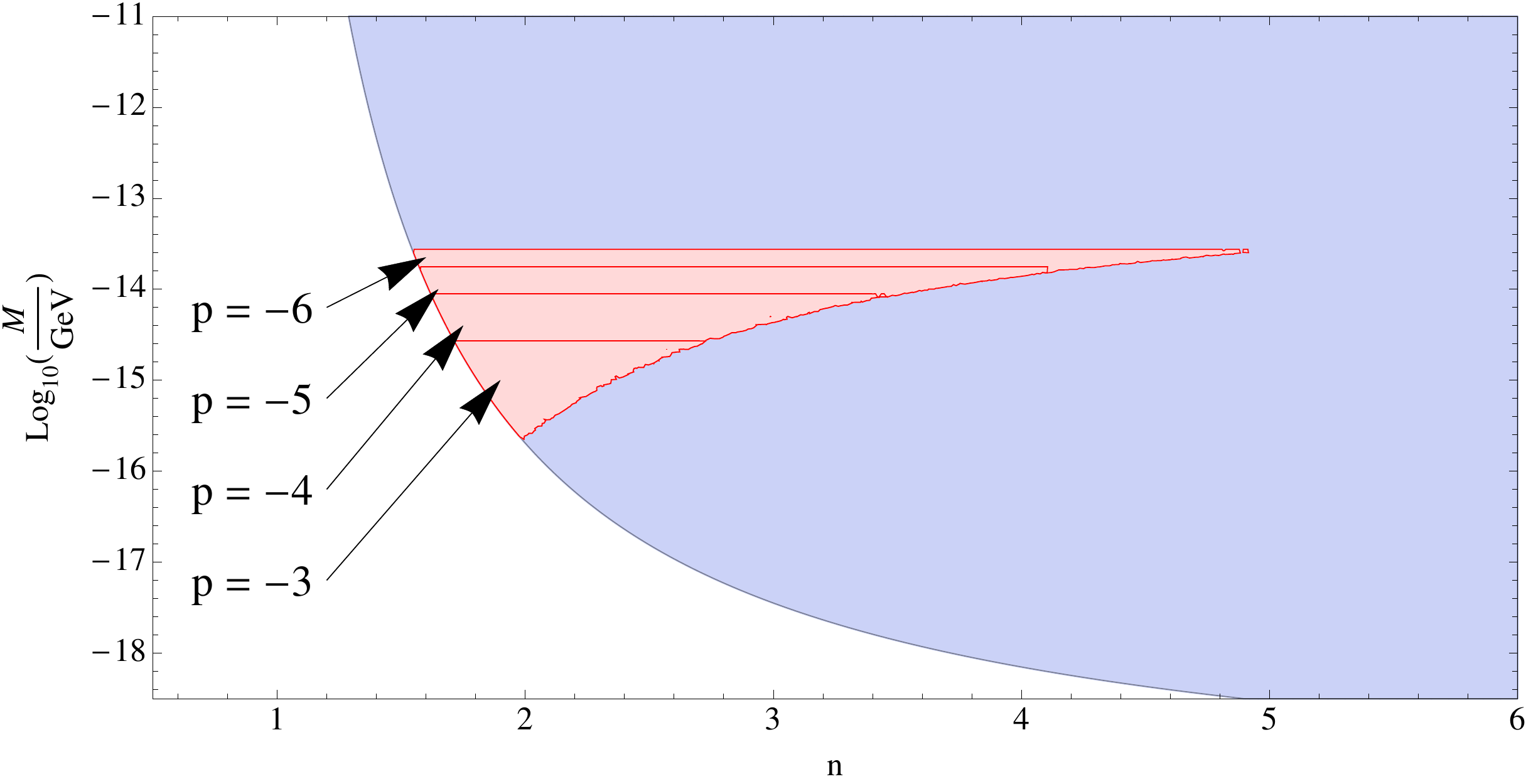}
\caption{Solar System constraints for model (\ref{constr_eps}) and (\ref{constr_phi}) with $M_1 = M_2 = M$ in the region $n\gtrsim 0.8$ when $p$ takes negative values (c.f.~Fig.~\ref{Fig:det_range_equal_mass}).}
\label{Fig:equal_mass_negative_p}
\end{figure}
For $p \lesssim -3$, in addition to the excluded zones shown in Fig.~\ref{Fig:det_range_equal_mass} for $n \lesssim 0.8$, there is also a small portion of the $n \gtrsim 0.8$ region which becomes forbidden by the experiments, in particular by Eq.~(\ref{SS_constr_3}) and, less strongly, by Eq.~(\ref{SS_constr_1}).
These regions have been drawn in Fig.~\ref{Fig:equal_mass_negative_p} for the values $p = -3,-4,-5,-6$.
Although they could not be presented in Fig.~\ref{Fig:det_range_equal_mass} simultaneously to the positive $n$ constraints, it should be clear that the regions of Fig~\ref{Fig:equal_mass_negative_p} must be added to the ones in the $n \lesssim 0.8$ part.
For example the total Solar System constraints on the $p=-3$ model are the one depicted in Fig.~\ref{Fig:det_range_equal_mass} for $n\lesssim 0.8$ plus the small region shown in Fig.~\ref{Fig:equal_mass_negative_p} for the $n \gtrsim 0.8$ part.

Finally note that for $n=1$ only high values of $M$ lie in the detectable range and unless $p$ is extremely small ($p \lesssim 0.01$) such values are always excluded by the observations.
For $n=2$ instead all values of $M$ belonging to the detectable range, roughly from the Planck scale down to $M \sim 10^{-16}$, are excluded if $p\gtrsim 1$, while part of the detectable range is still viable if $p\lesssim 1$, becoming the whole $M$ detectable range when $p \ll 1$.
In the interesting constant mass case $n=0$ instead we notice that models with $p\lesssim -1$ exclude the entire detectable range for $M$.
Looking at Fig.~\ref{Fig:det_range_equal_mass} this trend can roughly generalized as follow: for $n<0$ the whole detectable range of $M$ is always forbidden if $p\lesssim n-1$.

The extended chameleon models (\ref{power_law_minimum})--(\ref{power_law_mass}) with $M_1 = M_2 = M$ thus allow for viable generalisations of the standard chameleon framework where the exponent $n$ not only can take values outside the range $1<n<2$, but it can also be negative or zero.
This last case is of great interest because it implies that there exists a viable screening mechanism where the effective mass of the scalar field does not change.
Moreover in the $p\lesssim -1$ case such model can successfully be restricted by current Solar System experiments.

\subsubsection{Double mass scale models}

Finally we can turn to the general case $M_1 \neq M_2$.
Since now there are four free parameters in (\ref{power_law_minimum})--(\ref{power_law_mass}), it will not be possible to simultaneously show the constraints for all of them.
We must fix two of them in order to draw pictures in the parameter space of the remaining two.
For this reason in the following we will only focus on the cases $n=-2,-1,0,1,2$.
We will thus draw the detectable range in the $(M_1,M_2)$ parameter space for these values of $n$ and show the Solar System constraints for different integer values of $p$, from $-3$ to $3$.
The result of this analysis is summarized in Fig.~\ref{fig:double_mass} below (ignore the green regions in Fig.~\ref{fig:double_mass} for the moment, they will be treated in Sec.~\ref{sec:galaxy_constprof_2mass}).

First of all note how the detectable range (depicted in blue in Fig.~\ref{fig:double_mass}) depends only on the values of $M_2$, since this is the mass scale entering the scalar field mass, as pointed out above.
Because of this, all the detectable ranges plotted in Fig.~\ref{fig:double_mass} can be obtained by the intersection of the detectable range of Fig.~\ref{Fig:det_range_equal_mass} with the vertical lines at $n=-2,-1,0,1,2$ respectively.
On the other hand the detectable region in the $(M_1,M_2)$ parameter space for the case $M_1=M_2$ reduces to the dashed lines shown in Fig.~\ref{fig:double_mass}.
The white region in Fig.~\ref{fig:double_mass} indicates again parameter values for which the fifth force becomes Yukawa suppressed and thus impossible to detect within the Solar System.
Note that for $n \lesssim 0.8$ (including $n=0$) high values of $M_2$ are Yukawa suppressed, while low values can possibly be detected (and thus constrained) within the Solar System.
On the contrary for $n \gtrsim 0.8$ low values of $M_2$ are detectable and high values are suppressed, in agreement with Fig.~\ref{Fig:det_range_equal_mass}.
The boundary of the detectable range will of course depend on the value of $n$, however for large negative and positive values of $n$ this boundary tends to converge toward the value $M_2 \simeq 10^{-20}\, {\rm GeV}$, as one can realise looking again at Fig.~\ref{Fig:det_range_equal_mass}.

The regions excluded by Solar System experiments are drawn in red and as before the ones with higher absolute value of $p$ include the ones with lower absolute value.
For example for $p=3$ ($p=-3$) the whole region excluded by Solar System constraints is represented not only by the red zone denoted with $p=3$ ($p=-3$), but also by the ones denoted by $p=2$ ($p=-2$) and $p=1$ ($p=-1$).
This means that the higher the absolute value of $p$, the more of the parameter space is excluded by the observations, irrespectively of the value of $n$.
The dominant constraints in this case are again given by Eqs.~(\ref{SS_constr_1}) and (\ref{SS_constr_3}), except for the ``protuberances'' appearing for positive $p$'s in the $n=1$ and, more evidently, in the $n=2$ case of Fig.~\ref{fig:double_mass} which are solely determined by the Cassini bound (\ref{SS_constr_3}).

When $M_1 \neq M_2$ the full parameter space of the extended chameleon models (\ref{power_law_minimum})--(\ref{power_law_mass}) is available and, as a consequence, values of $M$ that were excluded in the case of equal mass scales, can now become viable when $M_1 \neq M_2$.
For example, for $n=2$ and $p=2$ the whole detectable range is forbidden in the equal mass case, as can be realized looking at the bottom left plot in Fig.~\ref{fig:double_mass}, where the dashed line is completely inside the zone excluded by the observations, or observing that in Fig.~\ref{Fig:det_range_equal_mass} the detectable range restricted to the $n=2$ vertical line is completely excluded by Solar System experiments.
However for a fixed value of $M_1$ which was excluded in the $M_1=M_2$ case, say $10^{-12} {\rm GeV}$, a different value of $M_2$, say $M_2 \lesssim 10^{-18}$ GeV, can now lead to a viable model within the detectable range.
This is of course due to the fact that one has more parameters, and thus more freedom, if the two mass scales $M_1$ and $M_2$ are uncorrelated.

On the whole Fig.~\ref{fig:double_mass} shows that for the extended chameleon model (\ref{power_law_minimum})--(\ref{power_law_mass}) a great part of the detectable range is still allowed by the present observations and in principle could be probed by future experiments.
The situation is somehow different from the one of standard chameleons where, as shown by Fig.~\ref{Fig:det_range_chameleon}, almost all the detectable range seems to be forbidden by observations.

\section{Extended chameleons at galactic scales}
\label{sec:galaxy}

In this section we will investigate the possible effects of the extended chameleon model (\ref{power_law_minimum})--(\ref{power_law_mass}) at galactic distances.

\subsection{Constant density profile}

Before considering more realistic galaxy profiles, we will again employ the useful constant density profile (\ref{eq:sol_sphere_in})--(\ref{eq:sol_sphere_out}) in order to compare the expectations at galactic distances with the Solar System constraints derived in the previous sections.
In other words we will first model a galaxy as a spherically symmetric distribution of constant matter density immersed in a  surrounding environment, assumed again to be characterized by a constant density. Typically this outside matter distribution either represents the cosmological vacuum or the galaxy cluster halo in which the galaxy is embedded.
This clearly constitutes an oversimplification of the problem, but it will be useful to understand what to expect outside the virial radius. 
In fact we consider the galaxy to have a central core of radius $R_{\rm gal}$ immersed in a lower density which is a good approximation outside the virial radius $R_{\rm vir}$ of the galactic halo. We will take this outside density to extend up to a large radius $R_{\rm halo} \gg R_{\rm vir}$ which can be considered as the size of the galaxy cluster to which the studied galaxy belongs. We will consider this model as a first approximation
for the galaxy embedded in its cluster and valid for distances $r\gtrsim R_{\rm vir}$.
What we want to compute is the deviation from the Newtonian potential outside the virial radius  in order to estimate the fifth force effects acting on objects outside the dark matter halo of the galaxy.

First we must exclude any effects outside the galactic detectable range defined by $m_{\rm halo} R_{\rm halo} \lesssim 1$ where the Yukawa suppression guarantees the screening of the scalar field implications.
Adopting the extended chameleon model (\ref{power_law_minimum})--(\ref{power_law_mass}) such requirement transforms into
\begin{equation}
	 M_2 \left( \frac{\rho_{\rm halo}}{M_2^3 M_{\rm Pl}} \right)^{\frac{n}{2}} R_{\rm halo} \lesssim 1 \,,
	 \label{gal_det_range}
\end{equation}
where we estimate the radius of the cluster dark matter halo as $R_{\rm halo} \sim 10\, {\rm Mpc} \sim 10^{39}\, {\rm GeV}^{-1}$ and its density as $\rho_{\rm halo} \sim 200\, \rho_0$ where $\rho_0 \simeq 4.4\times 10^{-50} {\rm GeV}^{4}$ is the cosmological critical density.

We can evaluate the deviation from Newtonian gravity in  the region $r \gg R_{\rm vir}$ where $R_{\rm vir} \sim 200\, {\rm kpc} \sim 10^{37}\, {\rm GeV}^{-1}$ is our estimated value for the virial radius of the galaxy.
Within such region the first order correction to the Newtonian potential is in fact given by Eq.~(\ref{eq:eps_general}), which now reads
\begin{multline}
	\left|\epsilon_{\rm gal}\right| \approx C_{\rm gal}^{\rm halo} \frac{|p|}{\Phi_{\rm gal}} \frac{M_1^2 M_2^2}{\rho_{\rm halo}} \left( \frac{\rho_{\rm halo}}{M^3_2 M_{\rm Pl}} \right)^n \\
	\times \left( \frac{\rho_{\rm halo}}{M^3_1 M_{\rm Pl}} \right)^{-2p} \left| \left( \frac{\rho_{\rm halo}}{\rho_{\rm gal}} \right)^{p} -1 \right| \,,
	\label{eq:eps_galaxy}
\end{multline}
where the density inside the galaxy can be estimated as $\rho_{\rm gal} \sim 10^{-41}\, {\rm GeV}^4$, implying a Newtonian potential at its boundaries as $\Phi_{\rm gal} \sim 10^{-6}$.
Here the factor $C_{\rm gal}^{\rm halo}$ is given by Eq.~(\ref{eq:const_out}) with $\rho_{\rm in} = \rho_{\rm gal}$, $\rho_{\rm out} = \rho_{\rm halo}$ and $R = R_{\rm gal}$.

\subsubsection{Standard chameleons}

\begin{figure}
\includegraphics[width=0.95\columnwidth]{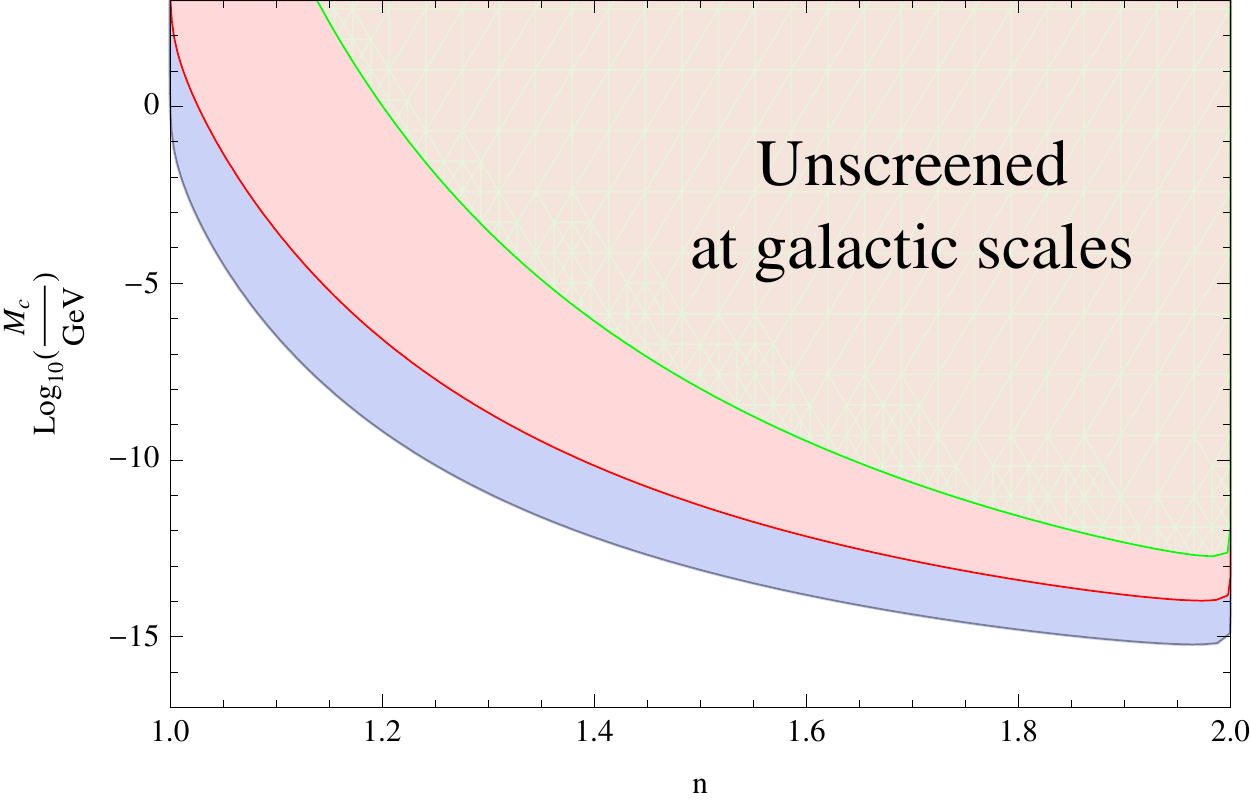}
\caption{Galactic screening for the chameleon model (\ref{minimum_chameleon})--(\ref{mass_chameleon}). The green area denotes the region where possible effects of the fifth force could be detected at galactic distances.
The rest of the picture coincides with Fig.~\ref{Fig:det_range_chameleon}.}
\label{Fig:galaxy_chameleon}
\end{figure}
\begin{figure*}
\includegraphics[width=0.9\textwidth]{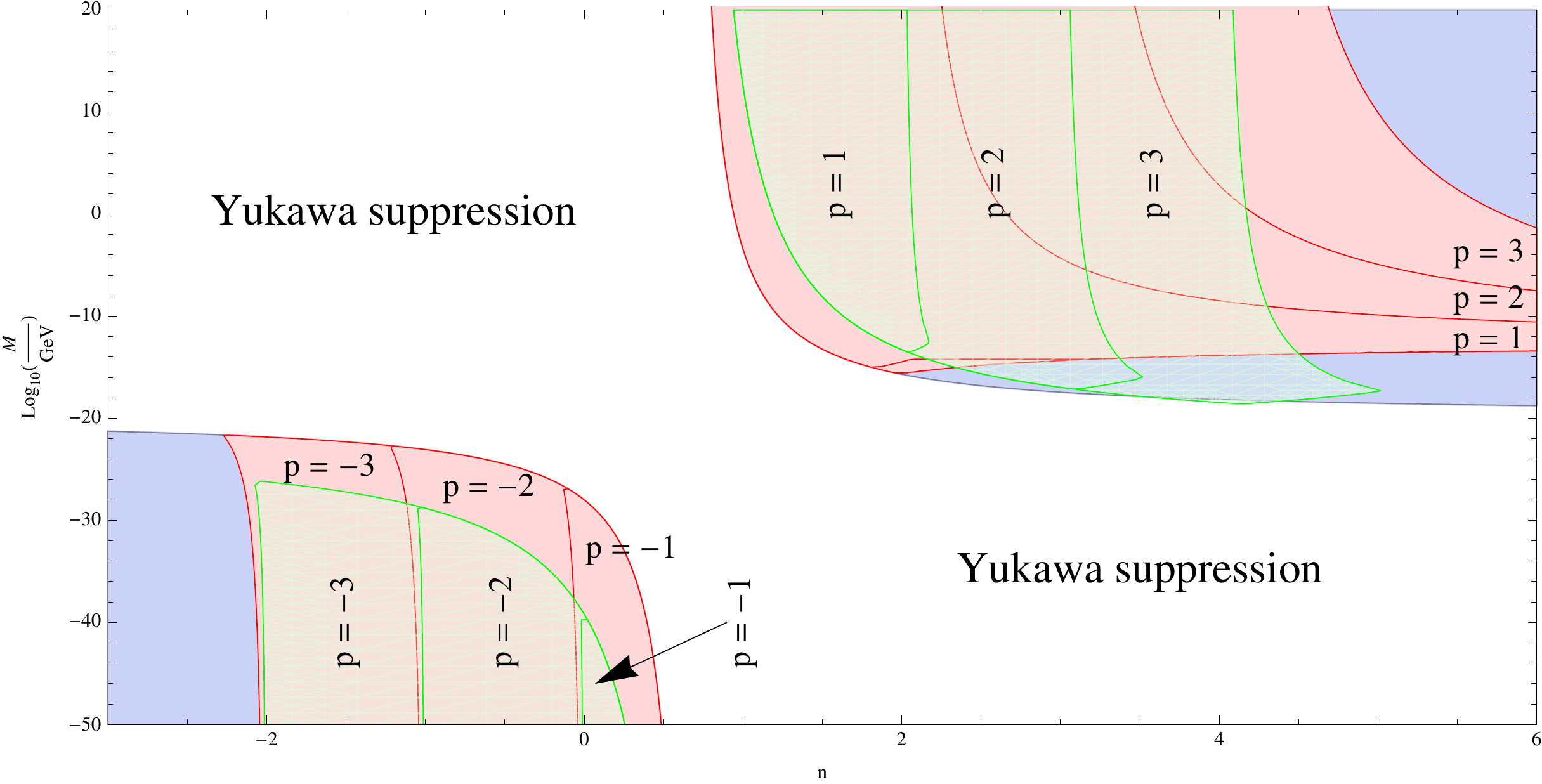}
\caption{Galactic screening for the models (\ref{power_law_minimum})--(\ref{power_law_mass}) with $M_1 = M_2 = M$. The green zones denotes the region of the parameter space where possible fifth force effects might be detected at galactic distances. They become larger as $\vert p\vert $ increases. The rest of the figure coincides with Fig.~\ref{Fig:det_range_equal_mass}.}
\label{Fig:galaxy_single_mass}
\end{figure*}
We can first analyse the standard chameleon case (\ref{minimum_chameleon})--(\ref{mass_chameleon}) in order to understand whether such well-known model admits possible applications at galactic scales.
In Fig.~\ref{Fig:galaxy_chameleon} we have plotted in green the region where $|\epsilon_{\rm gal}| > 0.01$, i.e.~where the fifth force is at least 1\% of the Newtonian force.
This should indicatively denote the region where fifth force effects can possibly be detected by present or future observations at galactic distances.
As  clear from Fig.~\ref{Fig:galaxy_chameleon}, for the standard chameleon model this region of the parameter space is wholly contained within the part excluded by Solar System experiments.
This implies that chameleon effects at galactic scales are either ruled out by present constraints or completely screened.
It seems thus improbable to find any fifth force effect within galaxies like the Milky Way for the standard chameleon model. On the other hand, models with the  chameleon mechanism such as $f(R)$ models in the large curvature limit could be tested using dwarf galaxies with a surface Newton potential below $10^{-7}$. In the following, we shall restrict ourselves to cases like the Milky Way where the standard chameleon do not yield interesting results on galactic scales and analyse models where this is no longer the case, i.e.~where the scalar field can play a non-negligible role either inside or outside the virial radius of such galaxies.

\subsubsection{Single mass scale models}

At this point we can repeat the analysis for the single mass scale models where the minimum (\ref{power_law_minimum}) and mass (\ref{power_law_mass}) are related by assuming $M_1 = M_2 = M$.
In Fig.~\ref{Fig:galaxy_single_mass} we have drawn for different values of $p$ the regions of the $(p, M)$ parameter space where $|\epsilon_{\rm gal}| > 0.01$ (depicted in green with vertical labels).
These should roughly indicate where fifth force effects at galactic scales might be detected by present or future experiments capable of probing the Newtonian force with a 1\% accuracy.
As one can realise looking at Fig.~\ref{Fig:galaxy_single_mass}, for negative values of $p$ these regions are completely contained within the area excluded by Solar System experiments and thus no deviations from Newton's law should appear outside the virial radius.
Note that this holds also in the interesting $n=0$ case, where the scalar field mass is constant.
For positive values of $p$ however there are regions of the parameter space where possible fifth force effects in the galaxies are not ruled out by Solar System constraints.
From Fig.~\ref{Fig:galaxy_single_mass} we see that this happens for $p= 2, 3$, but not for $p=1$ whose green region is completely immersed in the corresponding excluded red region.
Hence if $p\gtrsim 2$ there are combinations of the model parameters for which possible galactic deviations from Newtonian gravity might be present.
Interestingly a major part of these regions lies within the Solar System detectable range and could thus be excluded by future local experiments, unless galactic observations will manage to rule them out first.

\subsubsection{Double mass scale models}
\label{sec:galaxy_constprof_2mass}

\begin{figure*}
\includegraphics[width=.96\textwidth]{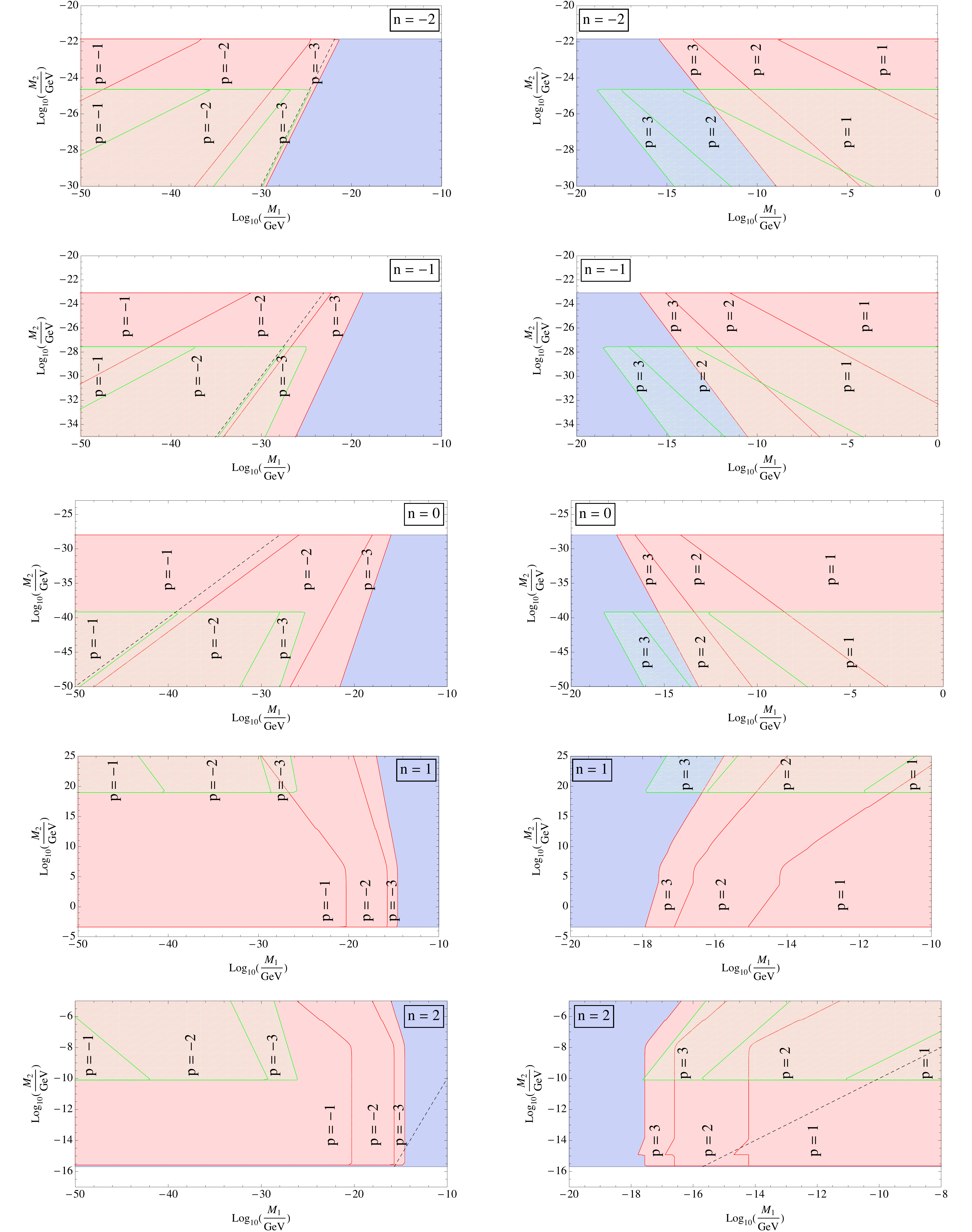}
\caption{Solar System constraints (red) and no galactic screening (green) for the models (\ref{power_law_minimum})--(\ref{power_law_mass}) with values of $n$ from -2 to 2 and values of $p$ from -3 to 3. In each panel the blue zone delimits the Solar System detectable range, while in the white region the fifth force is Yukawa suppressed.}
\label{fig:double_mass}
\end{figure*}

Finally it remains to investigate the full extended chameleon models (\ref{power_law_minimum})--(\ref{power_law_mass}) where the two mass scales are allowed to take different values.
The regions of the $(M_1, M_2)$ parameter space where $|\epsilon_{\rm gal}| > 0.01$ are shown in green in Fig.~\ref{fig:double_mass} for different values of $p$ from $-3$ to $3$.
Looking at the plots on the left hand side of Fig.~\ref{fig:double_mass}, it is clear that negative values of $p$ do not allow for possible fifth force effects at galactic distances (outside the virial radius) since they are always excluded by Solar System experiments.
On the other hand for positive values of $p$ there is always a region of the parameter space where deviations from Newton's force are big  on galactic scales and has not yet been excluded by local constraints.
The exception seems $n=2$ where only for $p \gtrsim 3$ the green area is not completely excluded by Solar System bounds.
However, even if not shown in Fig.~\ref{fig:double_mass}, for $n\gtrsim 3$ the Solar System observations fail to exclude the whole detectable region down to its boundary and part of it becomes again unscreened at galactic scales irrespectively of the value of $n$ (in other words even for small $n$s the green region is not completely excluded by the red one since it encompasses smaller values of $M_2$).
Note however that for $n=1$ the galactic unscreened regions appear only for $M_1 \gtrsim 10^{18}\, {\rm GeV}$. This may lead to  trans-Planckian theories, usually problematic from a quantum point of view.
In the interesting $n=0$ case however there is always a small part of the Solar System detectable range which is not excluded by the observations and where possible galactic fifth force effects might be of a relevant magnitude.
For example choosing $n=0$ and $p=1$, from Fig.~\ref{fig:double_mass} one can see that the mass scales $M_1 = 10^{-10}\,{\rm GeV}$ and $M_2 = 10^{-40}\,{\rm GeV}$ are expected to provide non negligible deviations to the Newtonian attraction outside the virial radius.
On the other hand, as we are going to see, the model $n=0$ and $p=-1$ gives interesting effects within the virial radius.

\subsection{NFW profile}
\label{sec:NFW}

The aim of this section is to give an example of the possible effects that extended chameleons might provide at galactic distances, in particular regarding galaxy rotation curves. This can be achieved only penetrating inside the virial radius and for such analysis we will be using a realistic mass profile.
We will not provide constraints on the model parameters here since the comparison with observational data at galactic scales would demand for a more detailed investigation which lies outside the scope of our work.

We will model the galactic density using the well-known NFW profile defined by
\begin{equation}
	\rho_{\rm NFW} = \frac{\rho_s}{r/R_s \left(1+r/R_s\right)^2} \,,
	\label{eq:NFW_profile}
\end{equation}
where $\rho_s$ and $R_s$ are two constants which will be roughly assumed to be
\begin{equation}
	\rho_s \sim 10^{-41}\, {\rm GeV}^4 \quad\text{and}\quad R_s \sim 10^{36}\, {\rm GeV}^{-1} \, (10\, {\rm kpc}) \,,
	\label{eq:NFW_const}
\end{equation}
in the following figures.

The Newton potential obtained from the profile (\ref{eq:NFW_profile}) is given by
\begin{equation}
 	\Psi_N (u) = -4 \pi  G \rho_s R_s^2 \frac{\log (u+1)}{u} \,,
 	\label{eq:Newton_potential_NFW}
 \end{equation}
where $u = r / R_s$ and a constant of integration has been fixed to avoid a divergence at the origin \footnote{The resulting Newtonian force cannot be made to vanish at $r=0$ due to the divergence of the NFW profile (\ref{eq:NFW_profile}). This issue is usually solved matching the NFW profile with a finite core near the centre. We will however ignore the problem since it is irrelevant in the following analysis.}.
The Newtonian acceleration (proportional to the Newtonian force) is then simply given by the derivative of (\ref{eq:Newton_potential_NFW}) with respect to $r$.

We would like  to find the fifth force potential for the NFW profile (\ref{eq:NFW_profile}).
Unfortunately for the most general extended chameleon model (\ref{power_law_minimum})--(\ref{power_law_mass}), as well as for standard chameleons (\ref{minimum_chameleon})--(\ref{mass_chameleon}), it is extremely difficult (if not impossible) to solve the scalar field equation (\ref{eq:KG_eq_spherical}) analytically.
Nevertheless some analytical solutions might be found given some specific values of $n$ and $p$ (usually integer numbers).
In what follows we will consider the case $n=0$ and $p=-1$ motivating it with the following reasons:
\begin{itemize}
	\item It is the simplest model where a non-zero fifth force arises ($m$ is constant and $\phi_0$ is linear in $\rho$) and there exists an analytical solution of Eq.~(\ref{eq:KG_eq_spherical});
	\item It is a conceptually interesting case where only the minimum of the effective potential presents an environmental dependence, while its mass remains constant (new kind of screening);
\end{itemize}
In this  case, the Klein-Gordon takes a simple form which is amenable to analytical approximations when $r\lesssim R_s$ and $r\gtrsim R_s$. The Klein-Gordon equation reads:
\be
\frac{d^2 \phi}{dr^2}+ \frac{2}{r} \frac{d\phi}{dr} +M_2^2 \phi= \frac{\beta_{\rm test}}{M_{\rm Pl}} \rho_{\rm NFW}(r) \,,
\label{eq:approx_KG_1}
\ee
where we have introduced (see discussion in Sec.~\ref{sec:conclusions})
\be
\beta_{\rm test} = \frac{M^2_2}{M^2_1} \, .
\ee
On galactic scales we can neglect the mass term in Eq.~\eqref{eq:approx_KG_1} and get the field profile
\be
\frac{d\phi}{dr}\sim \frac{\beta_{\rm test} R_s\rho_s}{2M_{\rm pl}} + \frac{C}{r^2} \,, \quad\text{for} \ r\lesssim R_s \,,
\ee
where $C$ is an integration constant. Similarly we have
\be
\frac{d\phi}{dr}\sim \frac{\beta_{\rm test} R_s^3\rho_s}{M_{\rm pl}} \frac{\ln \frac{r}{R_s}}{r^2} +\frac{D}{r^2} \,, \quad\text{for} \ r\gtrsim R_s \,,
\ee
where $D$ is another integration constant.
The acceleration due to the scalar field, obtained deriving \eqref{eq:5th_force_pot} with respect to $r$, is
\be
a_5= -\frac{\beta_{\rm test}}{M_{\rm Pl}}( \frac{d\phi}{dr}-\frac{1}{M_{\rm Pl} M_1^2} \frac{d\rho_{\rm NFW}}{dr}) \,.
\ee
The constant $C$ is chosen to remove the diverge of the acceleration at the origin and we have that the acceleration is constant at $r=0$ (exactly as the Newtonian acceleration). Taking $C=-\frac{R_s\rho_s}{M_{\rm Pl} M_1^2}= -\frac{\beta_{\rm test} R_s \rho_s}{M_2^2 M_{\rm Pl}}$
and matching the approximate solutions at $r=R_s$ we find that
\be
a_5=-\frac{\beta_{\rm test}^2 R_s\rho_s}{2M_{\rm pl}^2 } \,, \quad\text{for} \ r\lesssim R_s \,,
\ee
and
\begin{multline}
a_5 \sim -\frac{\beta_{\rm test}^2 R_s^3\rho_s}{M_{\rm pl}^2r^2 }\left( {\ln \frac{r}{R_s}}+\frac{3}{M_2^2 r^2} -\frac{1}{M_2^2 R_s^2}+\frac{1}{2}\right)\,, \\ \text{for} \ r\gtrsim R_s \,. \label{a5}
\end{multline}
The fifth force increases from infinity to its maximum value at $r={\cal{O}}(R_s)$ and decreases to a constant for $r\to 0$. This will be confirmed by the exact analytical solution and numerical integration.
The same approximation gives for the Newtonian potential and the Newtonian acceleration
\be
a_N\sim -\frac{ R_s\rho_s}{4M_{\rm pl}^2 }\,, \quad\text{for} \ r\lesssim R_s \,,
\ee
and
\be
a_N \sim \frac{ R_s^3\rho_s}{4M_{\rm pl}^2r^2 }\left( \ln \frac{r}{R_s}+1\right)\,, \quad\text{for} \ r\gtrsim R_s \,. \label{aN}
\ee
This provides an order of magnitude for the ratio of the scalar to Newton acceleration
\be
\vert \frac{a_5}{a_N}\vert \sim \frac{4\beta_{\rm test}^2}{M_2^2 R_s^2} \,,
\label{rat}
\ee
at its maximum around $R_s$. This ratio is much larger than the ratio $2\beta_{\rm test}^2$ for point particles. Far away we have
\be
\vert \frac{a_5}{a_N}\vert \sim {4\beta_{\rm test}^2} \,, \quad\text{for} \ r\gg R_s \,,
\ee
which will be small when $\beta_{\rm test}$ is small enough as required to have $\frac{4\beta_{\rm test}^2}{M_2^2 R_s^2}\sim 1$ for galactic effects when $M_2 R_s \lesssim 1$.
This will be confirmed numerically: in other words (\ref{rat}) is in the right ballpark. We see that the fifth force
is enhanced in a galactic environment when $M_2 \lesssim 10^{-36}$ GeV. We will also see that no effects on large scale structures is guaranteed when $M_2 \gtrsim H_0$. It is in this interval of values
that the model is the most interesting as its effects are most prominent on galactic scales whilst having little relevance on both Solar System and cosmological distances.

The previous approximations can be confirmed by an exact calculation.
The analytical solution of Eq.~(\ref{eq:KG_eq_spherical}) in the $n=0$ and $p=-1$ model can be written in terms of the {\it exponential integral function} $\operatorname{Ei}(x)$ defined as
\begin{equation}
	\operatorname{Ei}(x) = -\int_{-x}^{\infty} \frac{e^{-t}}{t} dt \,.
\end{equation}
The general solution takes then the following expression
\begin{multline}
	\phi(u) = \frac{c_1}{R_s} \frac{e^{-\gamma u}}{u} +\frac{c_2}{R_s} \frac{e^{\gamma u}}{u} \\
	+\frac{\gamma ^2 e^{\gamma  (u+1)}}{2 R_s \Omega u} \operatorname{Ei}[- \gamma (u+1) ]+\frac{\gamma ^2 e^{ -\gamma  (u+1)}}{2 R_s \Omega u} \operatorname{Ei}[\gamma (u+1) ] \,,
\end{multline}
where $c_1$ and $c_2$ are two constants of integration and we have defined two dimensionless parameters as follows
\begin{equation}
	\gamma = M_2 R_s \,, \quad\text{and}\quad \Omega = \frac{ M_1^2 M_{\rm Pl} }{ R_s \rho_s } \,.
\end{equation}
We can now compute the fifth force potential using the general expression (\ref{eq:5th_force_pot}) which yields
\begin{multline}
	\Psi_5(u) = \log \Bigg\{1-\frac{\gamma ^2 }{ \rho_s R_s^4 u \Omega ^2}
		\Bigg[ c_2 \Omega  e^{ \gamma  u}+c_1 \Omega e^{-\gamma u} -1 \\
		+\frac{\gamma ^2}{2} e^{\gamma  (u+1)} \operatorname{Ei}[- \gamma(u+1)]
		+\frac{\gamma ^2}{2} e^{-\gamma  (u+1)} \operatorname{Ei}[\gamma (u+1)] \Bigg]\Bigg\} \,.
	\label{eq:5th_force_potential_NFW}
\end{multline}
The acceleration due to the fifth force (proportional to the fifth force itself) is then given by the derivative of this potential with respect to $r$.
In order to fix the integration constants we require the fifth force to vanish at infinity and to be finite at the origin (it cannot be made to vanish for the same reason the Newtonian force is not zero at $r=0$).
The first requirement is satisfied simply assuming $c_2 = 0$, while the second one leads to
\begin{equation}
	c_1 = \frac{1}{\Omega }-\frac{e^{\gamma } \gamma ^2 \operatorname{Ei}(-\gamma )}{2 \Omega }-\frac{e^{-\gamma } \gamma ^2 \operatorname{Ei}(\gamma )}{2 \Omega } \,.
\end{equation}
The final expression of the fifth force acceleration is quite involved and will not be presented here.
The reader interested in it can find the result simply deriving Eq.~(\ref{eq:5th_force_potential_NFW}) with respect to $r$.

We are now interested in the galactic rotation curves arising from the combination of the Newtonian and fifth force effects.
For this reason we need to compute the tangential velocity of an unscreened object rotating around the galaxy.
This is given by balancing the Newtonian plus fifth force with the centripetal force as
\begin{equation}
	v_{\rm tg}(r) = \sqrt{r\left[a_N(r) + a_5(r)\right]} \,,
\end{equation}
where $a_N$ and $a_5$ are the Newtonian and fifth force accelerations, respectively given by the derivative of Eqs.~(\ref{eq:Newton_potential_NFW}) and (\ref{eq:5th_force_potential_NFW}) with respect to $r$.

Now we want to find when $v_{\rm tg}$ is substantially modified by the presence of the fifth force.
This happens roughly when $a_5$ is bigger or around the same order of magnitude of $a_N$.
If $a_5 \gg a_N$ then the resulting galactic curves will significantly differ from the Newtonian results and thus it is highly probable that observations will eventually excludes such situations, though only a comparison with astronomical data can actually verify such a statement.
If instead $a_5 \ll a_N$ then the effects of the scalar field will be screened in a NFW profile, and thus no signs of the scalar field will appear at galactic distances.
The interesting range is thus determined by the condition $a_5 \sim a_N$, where the fifth force might provide small modifications to Newton's law and thus present observational signature in the rotation curves.
As one might expect this requires some fine tuning of the masses $M_1$ and $M_2$, i.e.~of the only two parameters of our model.
In what follows we will briefly discuss the single mass scale model, and then analyse the two mass scale models presenting some interesting examples.

\subsubsection{Single mass scale models} 

In the reduced $M_1 = M_2 = M$ case the fifth force effects on the galaxy rotation curves are always negligible if $M \gtrsim 10^{-35}\,{\rm GeV}$, while they are predominant if roughly smaller than $10^{-36}\,{\rm GeV}$.
This can be understood noting that in this case $\beta_{\rm test}=1$ and Eq.~(\ref{rat}) gives that values of $M$ much smaller than $10^{-36}$ GeV lead to very large scalar effects.
In fact using the solution (\ref{eq:5th_force_potential_NFW}) we find numerically that the interesting range where $a_5 \sim a_N$ is roughly between $10^{-36}$ and $10^{-35}\,{\rm GeV}$.
Unfortunately such range is already excluded by Solar System observations, as one can easily realise looking at Fig.~\ref{Fig:det_range_equal_mass} for the $n=0$ and $p=-1$ case.
It is then expected that galactic constraints on $M$, at least when a NFW profile is considered, will be weaker than the already available Solar System constraints we derived in the previous sections.
In fact what we can roughly deduce from our analysis at galactic distances is that $M \gg 10^{-35}\,{\rm GeV}$, which is worse than $M \gtrsim 10^{-27}\,{\rm GeV}$ deduced from Solar System experiments.


\subsubsection{Double mass scale models} 
\label{sec:rot_curves}

\begin{figure*}
\includegraphics[width=\textwidth]{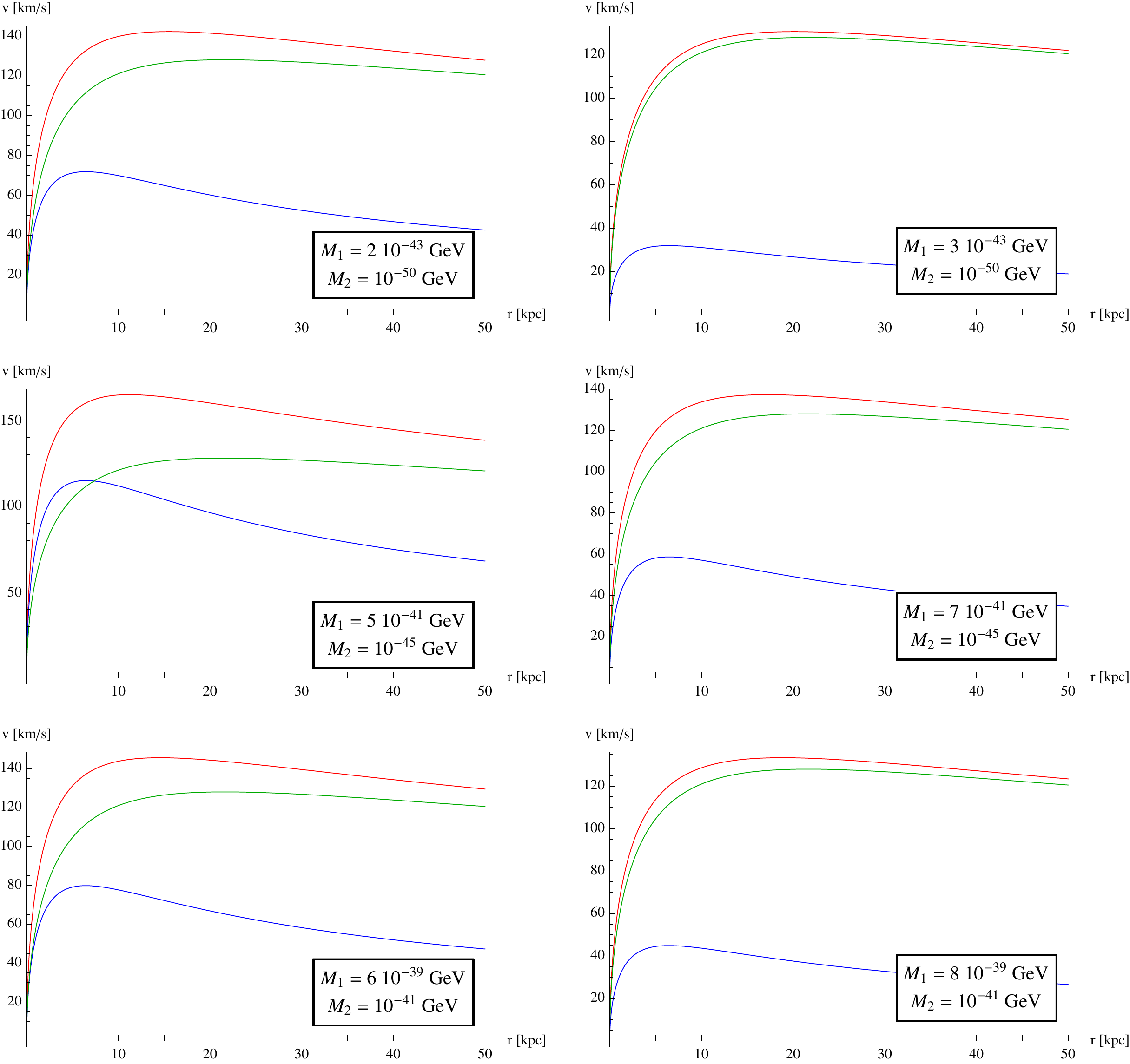}
\caption{Examples of galaxy rotation curves for the extended chameleon model (\ref{power_law_minimum})--(\ref{power_law_mass}) with $n=0$, $p=-1$ and different values of $M_1$ and $M_2$. The underlying matter distribution follows the NFW profile (\ref{eq:NFW_profile}) with $\rho_s$ and $R_s$ given in Eq.~(\ref{eq:NFW_const}). The blue line represents the contribution of the fifth force, the green line the contribution of the Newtonian force and the red line the total contribution of both forces combined.}
\label{fig:rot_curves}
\end{figure*}

In the full two mass scale model, we find more interesting results.
One can understand the situation looking at Fig.~\ref{fig:rot_curves}, where the galaxy rotation curves obtained for different values of $M_1$ and $M_2$ have been plotted.
To obtain such figures we have first chosen the value of $M_2$, taking into account the constraints of Fig.~\ref{fig:double_mass}, and then found values of $M_1$ for which $a_5 \sim a_N$.
Smaller values of $M_1$ and bigger values of $M_2$ will give rotation curves completely dominated by the fifth force and thus most likely to be excluded by experiments.
On the other hand, for bigger values of $M_1$ and smaller values of $M_2$ the effects of the fifth force will be negligible and the scalar field will be effectively screened in any galactic environment well described by a NFW profile.

\begin{figure}
\includegraphics[width=\columnwidth]{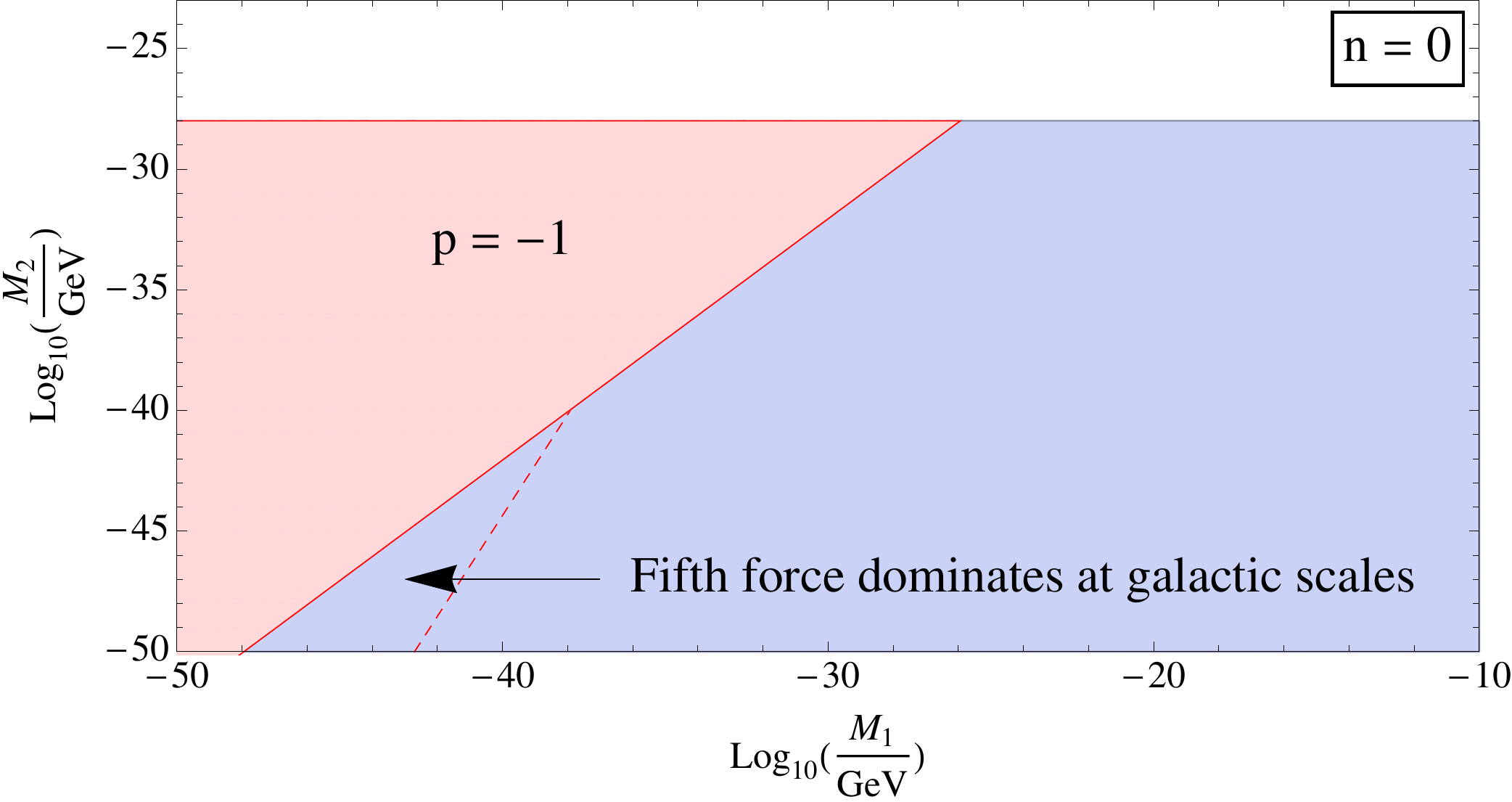}
\caption{The Solar System constraints exclude the red region when $n=0,\ p=-1$ while the effects of the scalar force would be of order larger than one at the core size $R_s$ of a NFW profile inside the small triangle region in blue.}
\label{fig:cmass_constr}
\end{figure}

This behaviour is summarized in Fig.~\ref{fig:cmass_constr}, where the centre-left panel of Fig.~\ref{fig:double_mass} has been reproduced only for the $p=-1$ case.
The range where $a_5 \sim a_N$, roughly determined by the red dashed line in Fig.~\ref{fig:cmass_constr}, now depends on both masses and for sufficiently small values of $M_2$ it happens to be not excluded by experiments and also to reside within the Solar System detectable range.
The examples in Fig.~\ref{fig:rot_curves} have also been chosen to show such features: to all values of $M_1$ and $M_2$ selected in the figures there corresponds a point in parameter space which lies close to the red dashed line.
Interestingly these points are all within the Solar System detectable range, but always outside and relatively close to the excluded region. 
This not only implies that the modifications of the rotation curves shown in Fig.~\ref{fig:rot_curves} are not yet ruled out by local constraints, but also that the screening mechanisms that produce them might be within the reach of future experiments.

Looking at Fig.~\ref{fig:rot_curves} we note that the fifth force contribution always decreases faster than the Newtonian one as $r\gtrsim R_s$.
In our approximate solution this can interpreted as coming from the fast decrease of the $1/(r^2M_2^2)$ term compared to the slow increase of $\ln r/R_s$ in (\ref{a5}) and (\ref{aN}).
This implies that at sufficiently large distances from the centre of the galaxy the scalar field will be effectively screened and only  Newton's law will determine the rotational motion of celestial objects.
On the other hand the deviations due to the fifth force is  significant around $R_s$ ($\sim 10$ kpc), which roughly coincides with the maximum of the Newtonian rotation curves, or even for smaller radii.
Interestingly such region is where the effects of the NFW profile on the observed galactic rotation curves are more prominent and thus deviations from the standard results might be tested more efficiently.
Note also that the fifth force for the NFW profile is always attractive since it leads to higher orbital velocities in the rotation curves.
A non negligible presence of the scalar field at galactic scales would thus imply that less dark matter would be needed in order to fit the current observations of the rotation curves.

Finally we notice that all the values of $M_1$ and $M_2$ in Fig.~\ref{fig:rot_curves} happen to be outside the ($n=0$ and $p=-1$) green region of Fig.~\ref{fig:double_mass}, i.e.~the effects outside the virial radius using the constant profile are negligible.
As a matter of fact using the NFW profile we find the same result for $r \gg R_s$ where the fifth force becomes completely screened and the results obtained with the two profiles, constant sphere and NFW, agree.

Although some dose of fine tuning on $M_1$ and $M_2$ is necessary in order to achieve the condition $a_5 \sim a_N$, i.e.~using (\ref{rat}) we must have $M_2 \sim M_1^2 R_s /2$, the effects of the fifth force on the galactic rotation curves presented in Fig.~\ref{fig:rot_curves} might constitute new possible signatures to look for in the astronomical observations.
For this reason the comparison of these results against the observational data of various galaxy rotation curves, which might provide stronger constraints on the two mass scales of the model or hints of the presence of the scalar field,  deserves further study.

\section{Extended Chameleons at Cosmological and Cluster Scales}
\label{sec:cosmo}

In this section we will finally analyse the new extended chameleon models at the largest scales of the universe.
We will focus on the constant mass model with $n=0$ and $p=-1$ since, as shown in the previous section, it can lead to interesting observational signatures at galactic scales.

\subsection{Background Evolution}

The Scalar-Fluid models considered in this work modify the background cosmology in a way which is similar to Scalar-Tensor theories.
The full cosmological equations, at both background and linear perturbations levels, have been derived in \cite{part1,Koivisto:2015qua}.
Defining by
\be
Q_0= \frac{\partial V_{\rm eff}}{\partial\phi} = m^2 (\rho) (\phi- \phi_0 (\rho)) \,,
\ee
the coupling between the scalar and matter, we have a non-conservation equation for Cold Dark Matter (CDM)
\be
\dot \rho + 3H \rho = Q_0 \dot \phi \,,
\ee
 while the Klein-Gordon equation becomes
 \be
 \ddot \phi + 3H \dot \phi  = -Q_0 \,,
 \ee
 where the effective potential of the cosmological model is taken to be
 \be
 V_{\rm eff} (\phi) = V_0 + \frac{1}{2} m^2(\rho)  (\phi- \phi_0(\rho))^2 \,,
 \label{eq:pot_cc}
 \ee
 that is to say we have introduced a constant energy density $V_0$ which is responsible for the later time acceleration of the expansion of the Universe (c.f.~Eq.~(\ref{eq:eff_square_pot})).
Note that in a neighbourhood of the minimum $\phi_0$ the potential (\ref{eq:pot_cc}) could well approximate any effective potential for the scalar field; for example $V_{\rm eff} = V_0 \cosh \left[m\left(\phi-\phi_0\right)/\sqrt{V_0}\right]$.

The full cosmological dynamics of these models will not be studied here.
We will focus on the $n=0, \ p=-1$ case as it has nice features on galactic scales (c.f.~Sec.~\ref{sec:NFW}).
In this case, as long as $m(\rho) = M_2 \gg H_0$, one can show that the minimum of the effective potential
 \be
 \phi_{\rm min}= \phi_0(\rho) \,,
 \ee
  is a cosmological tracker for the background cosmology. Indeed, putting
  \be
 \phi = \phi_{\rm min} + \delta\phi \,,
 \ee
  we have
\be
 \ddot{\delta \phi} + 3H \dot{\delta \phi} + m^2 (\rho) \delta \phi =  - \rho \frac{d}{dt} ( \frac{\dot \phi}{\rho}) \,,
 \ee
 where $\rho = \frac{\rho_0}{a^3}$ and $3M_{\rm Pl}^2 H^2\simeq \rho$ in the matter dominated era (when $\rho \gg \rho_\phi$).
Now in a first approximation at the minimum $\phi \simeq \rho/(M_1^2 M_{\rm Pl})$, implying $Q_0 \simeq 0$ and consequently $\dot\rho + 3 H \rho \simeq 0$. We thus can obtain $\dot \phi/\rho \simeq -3  H/ M_1^2 M_{\rm Pl}$ and finally
\be
\ddot \delta \phi + 3H \dot \delta \phi + m^2 (\rho) \delta \phi =  - \rho \frac { 3 H^2}{ 2 M_1^2 M_{\rm Pl}}.
 \ee
As long as $m (\rho) = M_2\gg H$, we have
\be
\delta\phi \sim   - \rho \frac { 3 H^2}{ 2 M_1^2 M_{\rm Pl} m^2(\rho)} \,,
\ee
 and therefore
 \be
 \vert \frac{\delta \phi}{\phi}\vert = \frac{ 3 H^2}{ 2 m^2 (\rho)}\ll 1 \,,
 \ee
 implying that the minimum is a tracker solution.
For instance with $M_2=10^{-41}\,{\rm GeV} \gg H_0$, the tracking solution is  valid at low $z$ in the late time Universe.

\subsection{Structure Formation}

Even though the background cosmology is identical to $\Lambda$-CDM as long as the field tracks the minimum of the potential, there could be effects on the growth of large scale structures. For the $n=0$, $p=-1$ (constant mass) model, the density contrast $\delta=\frac{\delta \rho}{\rho}$ satisfies (see \cite{Koivisto:2015qua} for more details)
\be
\ddot \delta + (2+ 3 c_s^2) H \dot\delta + (c_s^2 \frac{k^2}{a^2} -\frac{\rho}{2M_{\rm Pl}^2 }C) \delta=0 \,,
\ee
where the speed of sound is defined by
\be
c_s^2= \frac{M_2^2}{M_1^4 M_{\rm Pl}^2} \rho \,.
\ee
Notice that in the cases of interest where $H\ll M_2 \ll M_1$, we have that $c_s^2 \ll 1$ and the effects of the speed of sound can be neglected.
Similarly we have that
\begin{multline}
	C=1+ 36 c_s^2 \frac{H^2 \rho}{M_1^4 M_{\rm Pl}^2} -18c_s^2 \frac{ H^2 M_{\rm Pl}^2}{\rho} \\
	+6 (5 + 6 c_s^2) \frac{M_2^2}{M_1^2} \frac{H^2}{M_1^2} - 18 \frac{M_2^4}{M_1^4} \frac{H^2 \rho}{M_1^4 m_{\rm pl}^2} +2 \frac{M_2^4}{M_1^4} \,.
\end{multline}
Again, as long as $H\ll M_2 \ll M_1$, all the correction terms are small and we have $C \approx 1$,
implying that the growth of structure is not affected by the presence of a scalar degree of freedom and behaves like in the $\Lambda$-CDM model at late times.

Note that taking $H \sim H_0 \sim 10^{-42}\,{\rm GeV}$ the condition $H\ll M_2 \ll M_1$ is attained in the last examples (bottom panels) of Fig.~\ref{fig:rot_curves}, whilst it does not hold in the other cases.
This means that the constant mass model with $n=0$ and $p=-1$ can effectively screen the scalar field at Solar System and cluster scales while giving interesting results at galactic scales which might lead to possible observational signatures.

\section{Discussions \& Conclusions}
\label{sec:conclusions}

In this work we have employed the framework of Scalar-Fluid theories to introduce and investigate new kinds of couplings between a scalar field and the surrounding matter energy density.
In particular we have defined the effective scalar field interaction by a quadratic potential where the mass and minimum can arbitrarily depend on the matter density.
Thanks to its general dependence on the environment, this quadratic effective potential can in principle well approximate any scalar field potential, as long as only values in the neighbourhood of the minimum are considered.
Within this broad context we have found the conditions under which the fifth force due to the scalar field is screened within the Solar System and discussed its relation with the present experimental constraints.

An important result follows: if the minimum of the effective potential does not depend on the matter energy density, i.e.~it is a constant, then the fifth force effects are completely negligible.
This is true not only at the macroscopic level, which we have explicitly checked at both Solar System and galactic distances, but also at the microscopic level, since according to Eq.~(\ref{eq:test_particle_coupling}) the fifth force between two test particles always vanishes if the minimum is constant.
On the contrary,  a constant mass does not generally preclude the fifth force from providing detectable effects at all scales. This type of models differs drastically from chameleons, dilatons and symmetron where the mass is always varying with the density. We even find that such models can provide unexpected effects on galactic scales while preserving the solar system gravity and the growth of large scale structure.

We have shown that a subclass of the general Scalar-Fluid models, which we named {\it extended chameleons}, nicely generalises and includes the well known chameleon theories.
For these models we have derived constraints on the corresponding parameter spaces comparing their effects against Solar System observations.
Moreover we have studied the possible effects at galactic distances, first outside and then within the virial radius, where we have shown that interesting deviations with respect to the usual NFW profile might appear in the galaxy rotation curves.
These constitute possible observational signatures to look for in comparisons with astronomical data, which have been left for future work.
We have also checked that the dynamics at cosmological scales, at both background and linear perturbations levels, correspond with the one obtained with $\Lambda$CDM and that the growth of structure is unaffected by the presence of the scalar field, provided certain conditions are satisfied.

Within this class of extended chameleons we have found that a specific scalar field model stands out: the ones with a {\it constant mass} and a linearly dependent minimum.
These models  provide a new type of screening mechanism where only the minimum of the potential changes according to the environment, while the mass remains untouched.
For these models in Sec.~\ref{sec:rot_curves} we have obtained interesting effects on the galaxy rotation curves (see Fig.~\ref{fig:rot_curves}), which provide a two parameter modification of the common NFW results.
For some values of these two parameters the Solar System constraints are satisfied and the dynamics at cosmological scales become indistinguishable from $\Lambda$CDM, i.e. giving observable signatures at galactic distances, while not influencing  other scales.

Furthermore the constant mass (with linear minimum) model is also interesting  from a microscopic point of view.
Indeed the strength of the fifth force felt between two test particles is constant, as one can realise looking at Eq.~(\ref{eq:beta_test_EC}) which, for $n=0$ and $p=-1$, yields
\begin{equation}
	\beta_{\rm test} = \frac{M^2_2}{M^2_1} \,,
\end{equation}
where $M_1$ and $M_2$ are the two constant parameters of the model (respectively the mass scales of the minimum and the mass).
A constant $\beta_{\rm test}$ is obtained also with standard chameleon theories, while extended chameleons in general are characterized by a dependence of $\beta_{\rm test}$ on the matter energy density.
This suggests that the constant mass model could also be testable by laboratory experiments as the field profile would be very sensitive to the distribution of matter in Casimir or fifth force experiments.

In conclusion the results obtained in this work not only show that new consistent screening mechanisms for scalar fields can be defined employing the framework of Scalar-Fluid theories, but also that these new models naturally extend the well known chameleon theories and that they might provide interesting observational effects on the galaxy rotation curves, while passing all Solar System and cosmological tests.

\begin{acknowledgments}

We would like to thank P. Valageas for suggestions on the manuscript.
P.B.
acknowledges partial support from the European Union FP7 ITN
INVISIBLES (Marie Curie Actions, PITN- GA-2011- 289442) and from the Agence Nationale de la Recherche under contract ANR 2010
BLANC 0413 01. 
N.T.~acknowledge support from the Labex P2IO and an Enhanced Eurotalents Fellowship.

\end{acknowledgments}

\bibliography{ref2}

\end{document}